\documentclass[12pt]{article}
\usepackage{amssymb}
\usepackage{amsfonts}
\usepackage{myart}

\input tcilatex
\begin{document}

\title{General relativity extended to non-Riemannian space-time geometry.}
\author{Yuri A.Rylov}
\date{Institute for Problems in Mechanics, Russian Academy of Sciences,\\
101-1, Vernadskii Ave., Moscow, 119526, Russia.\\
e-mail: rylov@ipmnet.ru\\
Web site: {$http://rsfq1.physics.sunysb.edu/\symbol{126}rylov/yrylov.htm$}\\
or mirror Web site: {$http://gasdyn-ipm.ipmnet.ru/\symbol{126}%
rylov/yrylov.htm$}}
\maketitle

\begin{abstract}
The gravitation equations of the general relativity, written for Riemannian
space-time geometry, are extended to the case of arbitrary (non-Riemannian)
space-time geometry. The obtained equations are written in terms of the
world function in the coordinateless form. These equations determine
directly the world function, (but not only the metric tensor). As a result
the space-time geometry appears to be non-Rieamannian. Invariant form of the
obtained equations admits one to exclude influence of the coordinate system
on solutions of dynamic equations. Anybody, who trusts in the general
relativity, is to accept the extended general relativity, because \textit{%
the extended theory does not use any new hypotheses}. \textit{It corrects
only inconsequences and restrictions of the conventional conception of
general relativity. }The extended general relativity predicts an induced
antigravitation, which eliminates existence of black holes.
\end{abstract}

\section{Introduction}

In this paper we consider dynamic equations for the gravitational field
which are obtained at the generalization of the relativity theory on the
case of the most general space-time geometry. The general relativity
supposes, that the Riemannian geometry is the most general space-time
geometry. This supposition is based on our insufficient knowledge of a
geometry, when one supposes that any geometry is axiomatizable. It means,
that any geometry is constructed as a logical construction. In reality there
exist nonaxiomatizable space-time geometries \cite{R2001}, which are
constructed by means of the deformation principle \cite{R2007} as a
deformation of the proper Euclidean geometry. This geometry is described
completely by the world function \cite{S60} and only by the world function.
Such a geometry is called a physical geometry, because physicists need such
a geometry, which is a science on location of geometrical objects and on
their shape, (but not as a logical construction). Physicists use a geometry
as a tool for investigation of the space-time properties. Physicists are
indifferent to the question, whether or not a geometry is a logical
construction.

Physical geometry may be continuous, or discrete. It may be even granular,
i.e. partly continuous and partly discrete. The physical geometry is
described by the same manner in all cases. Properties of the physical
geometry are determined only by properties of the world function (but not by
properties of the point set, where the geometry is given). As a result the
physical geometry may be formulated in the coordinateless form (only in
terms of the world function). A good illustration of this fact is the
following example.

Let the proper Euclidean geometry be given in the Cartesian coordinates $%
\left( x,y\right) $ on the square $\left[ 0,1\right] \times \left[ 0,1\right]
$. It means, that the world function is given on this square. Let us map the
square $\left[ 0,1\right] \times \left[ 0,1\right] $ onto the
one-dimensional segment $\left[ 0,1\right] $. Let the mapping $\left(
x,y\right) \rightarrow X$ be one-to-one. It is possible only, if the mapping
is discontinuous at any point. For instance, the mapping can be realized as
follows. Let coordinates $x,$ $y$, $X$ be presented in the form of decimal
fractions%
\begin{equation}
x=0.\alpha _{1}\alpha _{2}\alpha _{3}....,\qquad y=0.\beta _{1}\beta
_{2}\beta _{3}....,\qquad X=0.\alpha _{1}\beta _{1}\alpha _{2}\beta
_{2}\alpha _{3}\beta _{3}...  \label{c1.0}
\end{equation}%
where $\alpha $ and $\beta $ are decimal ciphers. The formulas (\ref{c1.0})
realizes one-to-one mapping $\left( x,y\right) \leftrightarrow X$. Now the
world function $\sigma $ is given on one-dimensional segment $\left[ 0,1%
\right] $.%
\[
\sigma \left( X_{1};X_{2}\right) =\sigma \left(
x_{1},y_{1};x_{2},y_{2}\right) 
\]%
Nevertheless, considering the world function on the one-dimensional segment $%
\left[ 0,1\right] $, one can reconstruct the proper Euclidean geometry. In
particular, one can determine, that the geometry on the segment $\left[ 0,1%
\right] $ is the two-dimensional Euclidean geometry (in the sense, that the
maximal number of linear independent vectors is equal two), although the
geometry is given on one-dimensional segment.

For construction of a physical geometry it is sufficient to give a world
function for any pair of points of the point set, where the geometry is
given. One does not need to prove numerous theorems and to test a
compatibility of geometric axioms. The world function $\sigma \left(
P,Q\right) =\frac{1}{2}\rho ^{2}\left( P,Q\right) $ is a function of two
points$\ P$ and $Q$, where $\rho \left( P,Q\right) $ is a distance between
the two points. The number of possible world functions is much more, than
the number of infinitesimal intervals $dS^{2}=g_{ik}\left( x\right)
dx^{i}dx^{k}$, which are functions of only one point. In particular, there
is only one isotropic uniform geometry (the geometry of Minkowski) in the
set of Riemannian geometries. It is described by the world function $\sigma
_{\mathrm{M}}$. In the set of physical geometries any geometry is isotropic
and uniform, if it is described by the world function $\sigma =F\left(
\sigma _{\mathrm{M}}\right) $, where $F$ is an arbitrary function, having
the property $F\left( 0\right) =0$ and $\sigma _{\mathrm{M}}$ is the world
function of the geometry of Minkowski.

In particular, the space-time geometry, described by the world function%
\begin{equation}
\sigma =\sigma \left( \sigma _{\mathrm{M}}\right) =\sigma _{\mathrm{M}%
}+\lambda _{0}^{2}\text{sgn}\left( \sigma _{\mathrm{M}}\right) ,\qquad
\lambda _{0}^{2}=\frac{\hbar }{2bc}=\text{const}  \label{c1.1}
\end{equation}%
is uniform and isotropic. Here $\hbar $ is the quantum constant, $c$ is the
speed of the light and $b$ is some universal constant. Besides, this
geometry is nonaxiomatizable and discrete. In this space-time geometry any
motion of a free pointlike particles is multivariant (stochastic).
Statistical description of this stochastic motion is equivalent to the
quantum description in terms of the Schr\"{o}dinger equation \cite{R91}.
This circumstance admits one to obtain a statistical foundation of quantum
mechanics and to interpret quantum effects as geometrical effects. It admits
one to exclude the quantum principles from the set of prime physical
principles and to reduce the number of physical essences, what is important
for fundamental physical theories.

By definition the special relativity is a consideration of physical
phenomena in the flat uniform isotropic space-time. In the set of Riemannian
geometries there is only one such a geometry. It is the space-time geometry
of Minkowski. Description of physical phenomena in the geometry (\ref{c1.1})
should be qualified as an extended special relativity, because the
space-time geometry (\ref{c1.1}) is isotropic and uniform, but it is
non-Riemannian.

Statistical foundation of quantum theory shows also, that the real
space-time geometry may be non-Riemannian, and one cannot restrict oneself,
considering only the Riemannian space-time geometries.

Generalization of the general relativity on the case of physical space-time
geometry appears to be possible only at taking into account two essential
clauses:

\begin{enumerate}
\item Consideration of physical geometries.

\item Use of adequate relativistic concepts, and, in particular, a use of
the relativistic concept of the events nearness.
\end{enumerate}

The reasons of violation of the first condition are investigated in \cite%
{R2010}.

In the beginning of the twentieth century the theoretical physics developed
on the way of geometrization. The special relativity and the general
relativity were only stages of this geometrization. But the physics
geometrization appeared to be restricted by our poor knowledge of geometry,
when one knew only axiomatizable geometries. One was not able to work with
discrete geometries and geometries with restricted divisibility. Quantum
effects might be explained easily by multivariance of the space-time
geometry. However, the property of multivariance was not known in the
beginning of the twentieth century, and scientists were forced to introduce
new (quantum) principles of dynamics. As a result the quantum paradigm of
the microcosm physics development appeared. The quantum paradigm dominated
during the whole twentieth century. The quantum paradigm contains more
essences, than it is necessary.

In the end of the twentieth century, when our knowledge of geometry became
more complete, we may return to the program of further geometrization of
physics. The geometrical paradigm appeared to be possible, when, using
classical dynamic principles, quantum effects are freely explained by the
properties of the space-time geometry. The geometrical paradigm is more
attractive, because it uses less essences, than the quantum paradigm does.
To replace the quantum paradigm by the geometrical one, it is necessary to
generalize the special relativity and the general relativity on the case of
an arbitrary physical geometry of space-time.

The physical geometry is a geometry, which is described completely by the
world function, (world function is a half of the squared distance).
Practically, the physical geometry is a metric geometry, which is liberated
from all constraints on metric except the constraint, that the world
function (or metric) is equal to zero for two coinciding points. The
physical geometry is a very simple construction \cite{R2001,R2007}. For
constructing the physical geometry one does not need to formulate axioms and
to prove numerous theorems. It is sufficient to know the proper Euclidean
geometry, which is used as a standard physical geometry. All definitions of
the proper Euclidean geometry $\mathcal{G}_{\mathrm{E}}$ may be formulated
in terms of the Euclidean world function $\sigma _{\mathrm{E}}$. Replacing
the Euclidean world function $\sigma _{\mathrm{E}}$ in all definitions of
the Euclidean geometry by other world function $\sigma $, one obtains all
definitions of the physical geometry $\mathcal{G}$, described by the world
function $\sigma $.

Besides, the physical geometry is a monistic conception, which is described
by the only fundamental quantity $\sigma $. All other geometrical quantities
and concepts are expressed via fundamental quantity automatically. This
circumstance admits one to modify a physical geometry easily, because all
other geometric quantities concepts are modified automatically at
modification of the fundamental quantity $\sigma $. \cite{R2010a}. Program
of physics geometrization admits one to construct a monistic conception of
physics with the fundamental quantity $\sigma $.

The set of all Riemannian geometries is only a small part of the set of all
physical geometries. A generalization of the relativity theory on the case
of arbitrary physical geometry admits one to obtain such results, which
cannot be obtained in the framework of the Riemannian geometry. The
generalization of the special relativity (motion of particle in the given
space-time geometry) on the case of arbitrary physical space-time geometry
has been made already \cite{R2008}. A generalization of description of the
matter influence on the arbitrary space-time geometry met some problems.
These problems are connected with the fact, that in the relativity theory
some basic concepts are taken from the nonrelativistic physics. Concepts of
nonrelativistic physics are inadequate for consecutive geometric description
of the relativity theory and for generalization of this description on the
case of a more general space-time geometry.

Practically all physical phenomena on the Earth are nonrelativistic. At
first, we study nonrelativistic physics with its nonrelativistic concepts.
Relativistic effects appeared as corrections to nonrelativistic physics. In
the beginning of the twentieth century the relativistic physics was
presented in terms of slightly corrected nonrelativistic concepts. For
instance, the relativity principle has been presented as invariance of the
dynamic equations with respect to Lorentz transformation and as existence of
the supreme speed of the interaction propagation. Such a formulation is
useful for pedagogical goals, when one needs to transit from concepts of
nonrelativistic physics to relativistic ones. However, such a formulation is
not effective, when one tries to develop the relativistic physics. In this
case one should use concepts, which are adequate for relativistic physics.
In particular, the relativity principle is formulated in adequate concepts
as follows. The relativistic physics is a physics in the pseudo-Euclidean
space-time geometry of index 1 (geometry of Minkowski or that of
Kaluza-Klein). All other details of description are corollaries of
properties of the space-time geometry. For instance, properties, concerning
the role of the light speed, are pure geometrical properties of the
space-time.

Unfortunately, the formulation of the relativity theory in adequate
(geometrical) terms is used rare. The main difference of space-time geometry
in a relativistic theory from that in the non-relativistic (Newtonian)
physics is as follows. Relativistical event space (space-time) geometry is
described by one invariant (space-time interval), whereas in the Newtonian
physics the event space is described by two invariants (spatial distance and
temporal interval). Sometimes one does not mention this difference in
textbooks. Instead one speaks on difference in transformation laws. In
reality, the difference in the number of invariants is a fundamental
property, whereas the difference of the transformation laws is a very
special property, because it is essential only for flat space-time geometry.
Besides, the transformational properties are used only at description at
some coordinate system. They are useless at the coordinateless description.
This difference of formulations is not essential, when the theory is used
for calculation of concrete physical effects. However, this difference
becomes essential, if one tries to obtain a generalization of the relativity
theory on the case of the arbitrary space-time geometry.

For instance, the concept of a pointlike particle as a point in the
configuration space is a nonrelativistic concept. It needs concepts of
velocity and acceleration of this particle. These concepts are secondary
concepts, which can be introduced only after introduction of the linear
vector space and, in particular of a coordinate system. These concepts are
inadequate in the case of a discrete space-time geometry. As a result the
concept of velocity and that of acceleration cannot be used in an extension
of the relativity theory to the case of arbitrary space-time geometry, which
may be discrete.

In the general relativity all interactions (electromagnetic and
gravitational) are supposed to be short-range interactions. Concept of
short-range interaction is based on the nonrelativistic concept of the
events nearness. The events are considered as points in the event space
(space-time). Two events are considered to be near, if they happen in the
same place at the same time moment. This definition of nearness of events is
nonrelativistic, because this definition refers to a spatial distance and to
a temporal interval at once. A consistent relativistic concept of nearness
is to contain a reference to only quantity: space-time interval, (or world
function). For instance, if a supernew star flashed very far, and an
observer on the Earth observed this flash, the event of flash and the event
of this flash observation on the Earth are near (close) events.\label%
{0flash1}

According to common viewpoint the statement on nearness of the two events
(flash and observation of this flash) seems to be rather strange and
unexpected. However, from consistent relativistic viewpoint the two events
are near, because space-time interval between them is equal to zero.

The problem of relativistic concept of nearness is discussed in \cite{V2005}%
. It is known as the principle of Fokker \cite{F1929}, which is interpreted
as a conception of the action at a distance (but not as a relativistic
concept of nearness). The action at a distance is treated as a direct
influence of one object onto another one without intermediate agent
circulatory between them.

\section{Relativistic concept of nearness}

Let us consider the proper Euclidean geometry. Let $\rho \left( P,Q\right) $
be the Euclidean distance between the points $P$ and $Q$. The set $%
O_{\varepsilon }$ of points $P$, defined by the relation%
\begin{equation}
O_{\varepsilon }=\left\{ P|\ \rho \left( O,P\right) <\varepsilon \right\}
,\qquad \varepsilon >0  \label{a1.1}
\end{equation}%
is called $\varepsilon $-vicinity of the point $O$. If the parameter $%
\varepsilon $ is small, the points $P$ and $O$ are near \ $\left( P\simeq
O\right) $. If $\varepsilon \rightarrow 0$, $\varepsilon $-vicinity $%
O_{\varepsilon }$ degenerates into one point $O_{0}=O$. It is easy to see,
that, if $P\simeq Q$, then $Q\simeq P$.

The relation of nearness in the proper Euclidean geometry has the property
of transitivity: If $P\in O_{\varepsilon }$ and $Q\in O_{\varepsilon }$,
then $P\in Q_{2\varepsilon }$ and $Q\in P_{2\varepsilon }$. It follows from
the triangle axiom,%
\[
\rho \left( O,P\right) +\rho \left( O,Q\right) \geq \rho \left( Q,P\right) 
\]%
which is valid for the proper Euclidean geometry. If $\varepsilon $ $%
\rightarrow 0$, then $2\varepsilon $ $\rightarrow 0$ also. It means that, if 
$P\simeq O$ and $Q\simeq O$, then $P\simeq Q$.

The property of transitivity seems to be a natural property of the relation
of nearness. However, the transitivity property of the nearness relation
does not take place in the space-time geometry, for instance, in the
geometry of Minkowski. In this case the $\varepsilon $-vicinity $%
O_{\varepsilon }$ of the point $O$ is defined by the relation 
\begin{equation}
O_{\varepsilon }=\left\{ R|\ \left\vert \rho \left( O,R\right) \right\vert
<\varepsilon \right\} ,\qquad \rho \left( O,R\right) =\sqrt{2\sigma _{%
\mathrm{M}}\left( O,R\right) }  \label{a1.2}
\end{equation}%
Here $\sigma _{\mathrm{M}}\left( P,Q\right) =\sigma _{\mathrm{M}}\left(
x,x^{\prime }\right) $ is the world function of the space-time of Minkowski%
\begin{equation}
\sigma _{\mathrm{M}}\left( P,Q\right) =\sigma _{\mathrm{M}}\left(
x,x^{\prime }\right) =\frac{1}{2}\left( g_{\mathrm{M}}\right) _{ik}\left(
x^{i}-x^{\prime i}\right) \left( x^{k}-x^{\prime k}\right)  \label{a1.2a}
\end{equation}%
$x,$ $x^{\prime }$ are coordinates of points $P$ and $Q$ in some inertial
coordinate system, and $\left( g_{\mathrm{M}}\right) _{ik}$ is the metric
tensor in this coordinate system.

In this case the points with coordinates $P=\left\{ \sqrt{a^{2}+\varepsilon
^{2}},a,0,0\right\} $ and \newline
$Q=\left\{ \sqrt{a^{2}+\varepsilon ^{2}},-a,0,0\right\} $ belong to $%
\varepsilon $-vicinity of the point $O=\left\{ 0,0,0,0\right\} $, whereas $%
P\notin Q_{2\varepsilon }$, because%
\begin{equation}
\left\vert 2\sigma \left( P,Q\right) \right\vert =\left\vert \rho ^{2}\left(
P,Q\right) \right\vert =4a^{2}  \label{a1.3}
\end{equation}%
As far as the quantity $a$ may be indefinitely large, the distance between
the points $P$ and $Q$ may be very large, although both points are near to
the point $O$ ($P\simeq O$ and $Q\simeq O)$.

In the space-time of Minkowski the $\varepsilon $-vicinity of the point $%
O=\left\{ 0,0,0,0\right\} $ is a region of the space-time between two
hyperboloids%
\begin{equation}
\left( x^{0}\right) ^{2}-\mathbf{x}^{2}=\varepsilon ^{2},\qquad \left(
x^{0}\right) ^{2}-\mathbf{x}^{2}=-\varepsilon ^{2}  \label{a1.4}
\end{equation}%
Formally the relation (\ref{a1.4}) determines a sphere of radius $%
\varepsilon $ in the space-time of Minkowski. At $\varepsilon \rightarrow 0$
this region turns to the light cone with the vertex at the point $O$. Thus,
at $\varepsilon \rightarrow 0$ in the proper Euclidean geometry the $%
\varepsilon $-vicinity of the point $P$ is the same point $P$, whereas in
the geometry of Minkowski the $\varepsilon $-vicinity of the point $P$ is
the light cone with the vertex at the point $P$.

In the nonrelativistic physics the $\varepsilon $-vicinity $O_{\varepsilon }$
of the point $O=\left\{ 0,0,0,0\right\} $ is defined by relations%
\begin{equation}
O_{\varepsilon }=\left\{ \left\{ x^{0},\mathbf{x}\right\} |\ \left\vert
x^{0}\right\vert <\varepsilon \wedge \left\vert \mathbf{x}\right\vert
<\varepsilon \right\}  \label{a1.4a}
\end{equation}%
In the limit $\varepsilon \rightarrow 0$, the $\varepsilon $-vicinity (\ref%
{a1.4a}) turns to one point $O$. Thus, in the non-relativistic physics there
are only one near point, whereas in the relativistic physics there is a
continual set of near points. This difference appears to be very important
in definition of short-range interaction between particles.

Let us stress, that introducing cone-shaped $\varepsilon $-vicinity and
nearness of points on the light cone to the vertex of the cone, \textbf{we
do not suggest any hypothesis. We follow only the relativity principle.} If
we follow the relativity principles, we should accept the fact of the
cone-shaped $\varepsilon $-vicinity, because pointlike shape of the $%
\varepsilon $-vicinity in the limit $\varepsilon \rightarrow 0$ is a remnant
of the nonrelativistic theory.

\section{Relativistic concept of a pointlike particle}

In the consecutive geometric description any particle is realized by its
skeleton. In the case of a pointlike particle the skeleton is the ordered
set of two points $\left\{ P_{s},P_{s+1}\right\} $. The vector $\mathbf{P}%
_{s}\mathbf{P}_{s+1}$ describes the geometric momentum of particle. The
length $\left\vert \mathbf{P}_{s}\mathbf{P}_{s+1}\right\vert =\sqrt{2\sigma
\left( P_{s},P_{s+1}\right) }$ of the vector $\mathbf{P}_{s}\mathbf{P}_{s+1}$
describes the geometric mass of particle. Such a description is a pure
geometric one.

Motion of a pointlike particle is described by a world chain $\mathcal{C}$,
consisting of connected links $\mathcal{T}_{\left[ P_{s}P_{s+1}\right] }$ 
\begin{equation}
\mathcal{C=}\dsum\limits_{s}\mathcal{T}_{\left[ P_{s}P_{s+1}\right] }
\label{e1.1}
\end{equation}%
Any link $\mathcal{T}_{\left[ P_{s}P_{s+1}\right] }$ is a segment of
straight line, determined by the skeleton $\mathcal{P}_{1}^{\left( s\right)
}=\left\{ P_{s},P_{s+1}\right\} $. The link $\mathcal{T}_{\left[ P_{s}P_{s+1}%
\right] }$ is a set of points, determined by the relation 
\begin{equation}
\mathcal{T}_{\left[ P_{s}P_{s+1}\right] }=\left\{ R|\sqrt{2\sigma \left(
P_{s},R\right) }+\sqrt{2\sigma \left( R,P_{s+1}\right) }-\sqrt{2\sigma
\left( P_{s},P_{s+1}\right) }=0\right\}  \label{e1.2}
\end{equation}%
The length 
\begin{equation}
\mu =\sqrt{2\sigma \left( P_{s},P_{s+1}\right) }  \label{e1.3}
\end{equation}%
of all links is the same. The length $\mu $ is the geometric mass of the
particle, which is connected with the usual mass $m$ of the particle by the
relation 
\begin{equation}
m=b\mu =b\sqrt{2\sigma \left( P_{s},P_{s+1}\right) }  \label{e1.4}
\end{equation}%
where $b$ is the same universal constant, which appears in (\ref{c1.1})

Complicated (not pointlike) particles are described by a more complicated
skeleton $\mathcal{P}_{n}=\left\{ P_{0},P_{1},...P_{n}\right\} $ \cite{R2008}%
.

Description of the particle motion does not need an introduction of a
coordinate system. Details of such a description of the particle motion may
be found in \cite{R2008}. Such a description is generalized easily on the
case of arbitrary space-time geometry (in particular, discrete one). In the
microcosm the structure of the world chain (\ref{e1.1}) is essential, but
outside the microcosm one may consider the length $\mu $ of a link $\mathcal{%
T}_{\left[ P_{s}P_{s+1}\right] }$ to be infinitesimal, and to replace the
world chain by a smooth world line.

Let $\mathcal{L}$ be a world line of a pointlike particle, and the point $%
P\in \mathcal{L}$. A set $\mathbb{N}_{P}$ of events $Q$, which are near to
the point $P$ is different from the relativistic viewpoint and from the
nonrelativistic one. From nonrelativistic (conventional) viewpoint $\mathbb{N%
}_{P}=\left\{ P\right\} $, whereas from the relativistic viewpoint $\mathbb{N%
}_{P}=\mathcal{C}_{P}$, where $\mathcal{C}_{P}$ is the light cone with the
vertex at the point $P$. 
\begin{equation}
\mathcal{C}_{P}=\left\{ R|\sigma \left( P,R\right) =0\right\}  \label{a2.2}
\end{equation}

It is known, that the electromagnetic interaction between two pointlike
charged particles is carried out only via points, connecting with vanishing
space-time interval (retarding interaction), i.e. via points, which are near
from the relativistic viewpoint. The same is valid for the gravitational
interaction. On the other hand, the near points of the world line $\mathcal{L%
}$ should be interpreted in a sense of points belonging to the world line.
In this sense any interaction of two pointlike particles via near points may
be interpreted as a direct interaction (collision).

The light cones with vertexes at the points $P\in \mathcal{L}$, may be
considered as attributes of the pointlike particle, which is described by
the world line $\mathcal{L}$. We shall consider these light cones, directed
into the past, as bunches of isotropic straight lines $\mathcal{H}$. In
other words, any world line $\mathcal{L}$ of a pointlike particle is
equipped by bunches $\mathcal{C}_{P}$ of hair $\mathcal{H}_{P}$ at any point 
$P\in \mathcal{L}$. Any hair $\mathcal{H}_{P}$ consists of points $R\in 
\mathcal{H}_{P}$, which are near to the point $P\in \mathcal{L}$,$\quad $($%
\sigma \left( P,R\right) =0,\ \ \ R\in \mathcal{H}_{P}$) on the world line $%
\mathcal{L}$. The point $P$ is a footing of the hair $\mathcal{H}_{P}$. The
length of the hair $\mathcal{H}_{P}$ is equal to zero, because the hair $%
\mathcal{H}_{P}$ consists of points, which are near to the point $P$.
Although the length of any hair is equal to zero, nevertheless the hairs of
any world line cover the whole space-time. When some point $P^{\prime }\in 
\mathcal{H}_{P}$, $P\in \mathcal{L}_{1}$ of the world line $\mathcal{L}_{1}$
hair coincides with a point $P^{\prime }=P_{2}\in \mathcal{L}_{2}$ of other
world line $\mathcal{L}_{2}$, the particle $\mathcal{L}_{2}$ transfers a
part of its momentum to the particle $\mathcal{L}_{1}$. See figure

What part of its momentum does the particle $\mathcal{L}_{2}$ transfer,
depends on the point $P^{\prime }\in \mathcal{H}_{P}$, which is a common
point with $\mathcal{L}_{2}$ ($P^{\prime }=P_{2}\in \mathcal{L}_{2}$).

Although the length of any part of the hair is equal to zero, nevertheless
there is some invariant parameter along any hair $\mathcal{H}$. This
parameter $l_{r}$ is the relative length of the hair segment. The relative
length ($r$-length) of the point $P$ is the more, the "farther" the point $%
R\in \mathcal{H}_{P}$ lies from the footing $P$ of the hair $\mathcal{H}_{P}$%
. The $r$-length $l_{r}$ of the point $R\in \mathcal{H}_{P}$ is defined by
the relation 
\begin{equation}
l_{r}=l_{r}\left( P,R\right) =\frac{\left( \mathbf{PR.Q}_{0}\mathbf{Q}%
_{1}\right) }{\left\vert \mathbf{Q}_{0}\mathbf{Q}_{1}\right\vert }
\label{a2.3}
\end{equation}%
where vector $\mathbf{Q}_{0}\mathbf{Q}_{1}$ is an arbitrary timelike vector (%
$\sigma \left( O_{0},Q_{1}\right) >0$). The scalar product $\left( \mathbf{%
PR.Q}_{0}\mathbf{Q}_{1}\right) $ of vectors $\mathbf{PR}$\textbf{\ }and $%
\mathbf{Q}_{0}\mathbf{Q}_{1}$ is defined by the relation%
\begin{equation}
\left( \mathbf{PR.Q}_{0}\mathbf{Q}_{1}\right) =\sigma \left( P,Q_{1}\right)
+\sigma \left( R,Q_{0}\right) -\sigma \left( P,Q_{0}\right) -\sigma \left(
R,Q_{1}\right)  \label{a2.4}
\end{equation}%
\begin{equation}
\left\vert \mathbf{Q}_{0}\mathbf{Q}_{1}\right\vert =\sqrt{\left( \mathbf{Q}%
_{0}\mathbf{Q}_{1}\mathbf{.Q}_{0}\mathbf{Q}_{1}\right) }=\sqrt{2\sigma
\left( Q_{0},Q_{1}\right) }  \label{a2.5}
\end{equation}

It follows from expressions (\ref{a2.3}) - (\ref{a2.5}), that the relative
length is invariant, because it is expressed in terms of the world function.
Numerical value of the $r$-length depends on the choice of the timelike
vector $\mathbf{Q}_{0}\mathbf{Q}_{1}$. Sign of the $r$-length depends on the
choice of the timelike vector $\mathbf{Q}_{0}\mathbf{Q}_{1}$ also. However,
the order of points on the hair, directed to the past, is determined
single-valuedly by the value of the $r$-length.

If for some choice of the timelike vector $\mathbf{Q}_{0}\mathbf{Q}_{1}$ 
\begin{equation}
\left\vert l_{r}\left( P,R_{1}\right) \right\vert <\left\vert l_{r}\left(
P,R_{2}\right) \right\vert ,\qquad R_{1},R_{2}\in \mathcal{H}_{P}
\label{a2.6}
\end{equation}%
then the relation (\ref{a2.6}) takes place for any other choice of timelike
vector $\mathbf{Q}_{0}\mathbf{Q}_{1}$. It means that the point $R_{1}$ is
located between the points $P$ and $R_{2}$. The quantity of the transferred
momentum is inversely to the $r$-length $l_{r}\left( P,P^{\prime }\right) $
between the footing of the hair $P\in \mathcal{L}_{1}$ and the point $%
P^{\prime }\in \mathcal{L}_{2}$, $P^{\prime }\in \mathcal{H}_{P}$.

The concept of the world line hair admits one to consider and to calculate
electromagnetic and gravitational interaction of particles as a direct
collision of one particle with a hair of other particle. As far as the hairs
of a world line are considered as attributes of a particle, one may consider
electromagnetic and gravitational interaction of particles as a direct
collision of particles. Such a description of the particle interaction 
\textit{does not mention about gravitational and electromagnetic fields}. 
\textit{Such a description is a consecutive relativistic description. }

In the nonrelativistic theory the electromagnetic and gravitational fields
are \textit{essences, which exist independently of the matter}. These
essences provide the momentum transfer from one particle to another one.
Introduction of such essences is necessary, because the nonrelativistic
concept of nearness is used. In the consecutive relativistic theory, which
uses relativistic concept of nearness, one does not need to consider the
electromagnetic and gravitational fields as additional essences. It is
sufficient to consider them as a manner of description of particle
interaction. The less number of essences is contained in a fundamental
theory, the more perfect fundamental theory takes place.

Our conclusion, that gravitational and electromagnetic fields are not
physical essences (they are only attributes of the world function) seems
rather unexpected for most physicists. It is connected with the fact, that
the relativity theory is considered usually as a correction to the
nonrelativistic physics. As a result the relativity theory is presented
almost in all textbooks in terms of concepts of nonrelativistic physics. The
relativity theory is studied after presentation of nonrelativistic physics.
It is natural, that the relativity theory is presented in terms of
nonrelativistic concepts. Such a presentation is clearer for physicists,
which know nonrelativistic physics. New specific relativistic concepts are
used only in the case, when one cannot ignore them.

However, the relativity theory is a self-sufficient fundamental theory,
which can and must be presented without a mention of nonrelativistic
concepts. \textit{Furthermore, the relativity theory can be developed
successfully only in terms of adequate (relativistic) concepts}.

Let there be two timelike world lines $\mathcal{L}_{1}$ and $\mathcal{L}_{2}$
of two different particles. Any point $P\in \mathcal{L}_{1}$ corresponds, at
least, to one near point $P^{\prime }\in \mathcal{L}_{2}$, i.e. $P^{\prime
}\simeq P$, because the timelike world line $\mathcal{L}_{2}$ crosses the
light cone with the vertex at the point $P\in \mathcal{L}_{1}$. In other
words, any point of the world line $\mathcal{L}_{1}$ has a near point on the
world line $\mathcal{L}_{2}$ and vice versa.

Let us consider the space-time $\Omega $ of Minkowski, which is described by
the world function $\sigma _{\mathrm{M}}$, defined by (\ref{a1.2a}). Let the
inertial coordinate system $K$ be used, and the world chains $\mathcal{C}%
_{1},\mathcal{C}_{2}$ be timelike. The world chains $\mathcal{C}_{1}$ and $%
\mathcal{C}_{2}$ consist of connected segments $\mathcal{T}_{\left[
P_{l}P_{l+1}\right] }$ and $\mathcal{T}_{\left[ P_{l}^{\prime
}P_{l+1}^{\prime }\right] }$\label{0lengh}%
\begin{equation}
\mathcal{C}_{1}=\dbigcup\limits_{l}\mathcal{T}_{\left[ P_{l}P_{l+1}\right]
},\qquad \mathcal{C}_{2}=\dbigcup\limits_{l}\mathcal{T}_{\left[
P_{l}^{\prime }P_{l+1}^{\prime }\right] }  \label{f1.1}
\end{equation}%
\begin{equation}
\mathcal{T}_{\left[ P_{l}P_{l+1}\right] }=\left\{ R|\sqrt{2\sigma _{\mathrm{M%
}}\left( P_{l},R\right) }+\sqrt{2\sigma _{\mathrm{M}}\left( P_{l+1},R\right) 
}=\sqrt{2\sigma _{\mathrm{M}}\left( P_{l},P_{l+1}\right) }\right\}
\label{f1.2}
\end{equation}%
\begin{equation}
\mathcal{T}_{\left[ P_{l}^{\prime }P_{l+1}^{\prime }\right] }=\left\{ R|%
\sqrt{2\sigma _{\mathrm{M}}\left( P_{l}^{\prime },R\right) }+\sqrt{2\sigma _{%
\mathrm{M}}\left( P_{l+1}^{\prime },R\right) }=\sqrt{2\sigma _{\mathrm{M}%
}\left( P_{l}^{\prime },P_{l+1}^{\prime }\right) }\right\}  \label{f1.3}
\end{equation}%
All segments of a world chain have the same geometrical length $\mu $,
defined by the relation (\ref{e1.3}) The real mass of the particle,
described by the world chain, is connected with the geometric mass $\mu $ by
means of the relation (\ref{e1.4}).

Outside the microcosm the length $\mu $ is small with respect to
characteristic size of the world chain, and one may consider, that the
vectors $\mathbf{P}_{l}\mathbf{P}_{l+1}$ of any link have infinitesimal
length. In the Minkowski space-time $\Omega $ the timelike links $\mathcal{T}%
_{\left[ P_{l}P_{l+1}\right] }$ are one-dimensional infinitesimal timelike
segments. The timelike world chain $\mathcal{C}$ can be replaced by a smooth
timelike world line $\mathcal{L}$, whose points are labelled by a parameter $%
\tau $. The world line is described by the vector $\mathbf{PP}^{\prime
}\left( \tau \right) $, where $P$ is the origin of the coordinate system $K$
and $P^{\prime }\left( \tau \right) \in \mathcal{L}$. The vectors $\mathbf{P}%
_{l}\mathbf{P}_{l+1}$ of links turn into infinitesimal vectors $\mathbf{P}%
^{\prime }\left( \tau \right) \mathbf{P}^{\prime }\left( \tau +d\tau \right) 
$, which are tangent to the world line.

Let the world lines $\mathcal{L}_{1}$ and $\mathcal{L}_{2}$ be timelike. For
timelike world lines the infinitesimal segments $\mathcal{T}_{\left[
P_{l}P_{l+1}\right] }=$ $\mathbf{P}^{\prime }\left( \tau \right) \mathbf{P}%
^{\prime }\left( \tau +d\tau \right) $ are timelike, and the geometrical
mass $\mu $ is real ($\sigma \left( P_{l},P_{l+1}\right) >0$). In this case
the world lines $\mathcal{L}_{1}$ and $\mathcal{L}_{2}$ are one-dimensional,
and all points of a world line can be labelled by a parameter $\tau $.

As far as the space-time $\Omega $ of Minkowski is a linear vector space,
the vectors $\mathbf{PP}^{\prime }\left( \tau \right) $ can be represented
as a linear combination of basic vectors 
\begin{equation}
\mathbf{PP}^{\prime }\left( \tau \right) =f^{k}\left( \tau \right) \mathbf{e}%
_{k},  \label{f1.6}
\end{equation}%
where $\mathbf{e}_{k}$ are basic vectors of the inertial coordinate system $%
K $ with the origin $P$. The functions $f^{k}\left( \tau \right) $, $%
k=0,1,2,3$ are coordinates of points of the world line $\mathcal{L}_{2}$.

\label{0begin}Four basic vectors $\mathbf{e}_{k}$ may be presented in the
form 
\begin{equation}
\mathbf{e}_{k}=\mathbf{PQ}_{k},\qquad \mathbf{e}^{i}=\left( g_{\mathrm{M}%
}\right) ^{ik}\mathbf{e}_{k}=\left( g_{\mathrm{M}}\right) ^{ik}\mathbf{PQ}%
_{k},\qquad k=0,1,2,3  \label{f1.4b}
\end{equation}%
Here and further a summation over repeating Latin indices is produced $0\div
3$. The basic vector $\mathbf{e}_{k}=\mathbf{PQ}_{k}$ is determined by the
origin point $P$ and by the end point $Q_{k}$. Such a representation is
necessary to use the scalar product in arbitrary physical geometry, where
there is no linear space, and the scalar product of two vectors $\mathbf{PR}$
and $\mathbf{Q}_{0}\mathbf{Q}_{1}$ is defined by the relation (\ref{a2.4}).
The scalar product $\left( \mathbf{PR}.\mathbf{Q}_{0}\mathbf{Q}_{1}\right) $
of two vectors $\mathbf{PR}$ and $\mathbf{Q}_{0}\mathbf{Q}_{1}$ is defined
only via the world function without a reference to the properties of the
linear vector space.

Coordinates of the points $P^{\prime }\left( \tau \right) $ in the
coordinate system \ $K$ can be presented as follows 
\begin{equation}
\mathcal{L}_{2}:P^{\prime }\left( \tau \right) =\left\{ f^{k}\left( \tau
\right) \right\} =\left\{ \left( \mathbf{PP}^{\prime }\left( \tau \right) .%
\mathbf{e}^{k}\right) \right\} =\left\{ \left( g_{\mathrm{M}}\right)
^{ik}\left( \mathbf{PP}^{\prime }\left( \tau \right) .\mathbf{e}_{i}\right)
\right\} ,\ \ \tau \in \mathbb{R},\ \ P^{\prime }\in \Omega   \label{f1.5}
\end{equation}%
or%
\begin{equation}
f^{k}\left( \tau \right) =\left( g_{\mathrm{M}}\right) ^{kl}\left( \mathbf{PP%
}^{\prime }\left( \tau \right) .\mathbf{e}_{l}\right) =\left( g_{\mathrm{M}%
}\right) ^{kl}\left( \mathbf{PP}^{\prime }\left( \tau \right) .\mathbf{PQ}%
_{l}\right)   \label{f1.8}
\end{equation}%
\begin{equation}
f_{k}\left( \tau \right) =\left( g_{\mathrm{M}}\right) _{kl}f^{l}\left( \tau
\right) =\left( \mathbf{PP}^{\prime }\left( \tau \right) .\mathbf{PQ}%
_{k}\right)   \label{f1.8a}
\end{equation}%
where $\left( g_{\mathrm{M}}\right) ^{kl}$ is the contravariant metric
tensor, which is obtained from the covariant metric tensor $\left( g_{%
\mathrm{M}}\right) _{kl}$ by means of relations 
\begin{equation}
\left( g_{\mathrm{M}}\right) ^{il}\left( g_{\mathrm{M}}\right) _{lk}=\delta
_{k}^{i},\qquad \left( g_{\mathrm{M}}\right) _{lk}=\left( \mathbf{e}_{i}.%
\mathbf{e}_{k}\right) =\left( \mathbf{PQ}_{i}.\mathbf{PQ}_{k}\right) 
\label{f1.9}
\end{equation}%
In reality the functions $f^{k}\left( \tau \right) $ are piecewise. But for
simplicity we shall consider them as continuous and differentiable%
\begin{equation}
\mathcal{L}_{2}:\qquad x^{k}=f^{k}\left( \tau \right) ,\qquad \dot{f}%
^{k}\left( \tau \right) \equiv \frac{df^{k}\left( \tau \right) }{d\tau }%
\qquad k=0,1,2,3,  \label{f1.11}
\end{equation}

\section{Dynamic equations for calculation of world \newline
function of space-time}

Variation $\delta g_{ik}$ of the metric tensor, which is generated by
particles in the space-time geometry of Minkowski is described by the
relation \cite{F55}%
\begin{equation}
\left( c^{-2}\partial _{0}^{2}-\nabla ^{2}\right) \delta g_{ik}=-\kappa
T_{ik}  \label{a3.1}
\end{equation}%
where $T_{ik}$ is the energy-momentum tensor of particles. The constant $%
\kappa =8\pi G/c^{2}$, where $G$ is the gravitational constant and $c$ is
the speed of the light. Solution of this equation can be presented in the
form%
\begin{equation}
\delta g_{ik}\left( x\right) =-\kappa \int G_{\mathrm{ret}}\left(
x,x^{\prime }\right) T_{ik}\left( x^{\prime }\right) \sqrt{-g_{\mathrm{M}}}%
d^{4}x^{\prime },  \label{a3.2}
\end{equation}%
\begin{equation}
g_{\mathrm{M}}=\det \left\vert \left\vert \left( g_{\mathrm{M}}\right)
_{ik}\right\vert \right\vert ,\qquad i,k=0,1,2,3  \label{a3.2a}
\end{equation}%
where the retarded Green function $G_{\mathrm{ret}}\left( x,x^{\prime
}\right) $ has the form 
\begin{equation}
G_{\mathrm{ret}}\left( x,x^{\prime }\right) =\frac{1}{2\pi c}\theta \left(
x^{0}-x^{0\prime }\right) \delta \left( 2\sigma _{\mathrm{M}}\left(
x,x^{\prime }\right) \right)   \label{a3.3}
\end{equation}%
Here $\sigma _{\mathrm{M}}$ is the world function of the Minkowski
space-time, defined by the relation\ (\ref{a1.2a}), and the multiplier 
\begin{equation}
\theta \left( x\right) =\left\{ 
\begin{array}{ccc}
1 & \text{if} & x>0 \\ 
0 & \text{if} & x\leq 0%
\end{array}%
\right.   \label{a3.6}
\end{equation}%
The energy-momentum tensor $T_{ik}$ of a particles has the form 
\begin{equation}
T^{ik}\left( x\right) =\dsum\limits_{s}p_{\left( s\right) }^{i}\left(
x\right) u_{\left( s\right) }^{k}\left( x\right) ,\qquad i,k=0,1,2,3
\label{a3.7}
\end{equation}%
where $u_{\left( s\right) }^{k}\left( x\right) $ and $p_{\left( s\right)
}^{k}\left( x\right) $ are distributions of the 4-velocity and of the
4-momentum of the $s$th particle in the space-time. We have for the particle
number $s$%
\begin{equation}
\mathcal{L}_{\left( s\right) }:\qquad x^{i}=f_{\left( s\right) }^{i}\left(
\tau \right) ,\qquad p_{\left( s\right) }^{i}=b\dot{f}_{\left( s\right)
}^{i}\left( \tau \right) d\tau ,\qquad g_{\mathrm{M}}^{ik}p_{\left( s\right)
i}p_{\left( s\right) k}=m_{\left( s\right) }^{2}c^{2},\qquad s=..0,1,...
\label{a3.8}
\end{equation}%
\begin{equation}
u_{\left( s\right) }^{k}=\frac{\dot{f}^{k}\left( \tau \right) }{\sqrt{g_{rs}%
\dot{f}_{\left( s\right) }^{r}\left( \tau \right) \dot{f}_{\left( s\right)
}^{s}\left( \tau \right) }}  \label{a3.8a}
\end{equation}%
\begin{equation}
p_{\left( s\right) i}=g_{\mathrm{M}ik}b\left( f_{\left( s\right) }^{k}\left(
\tau +d\tau \right) -f_{\left( s\right) }^{k}\left( \tau \right) \right) =g_{%
\mathrm{M}ik}\dot{f}_{\left( s\right) }^{k}\left( \tau \right) bd\tau 
\label{a3.9}
\end{equation}%
where the constant $b$ is the proportionality coefficient (\ref{e1.4})
between the length of the world line link $\mu =\left\vert \mathbf{P}_{l}%
\mathbf{P}_{l+1}\right\vert $ and the particle mass, described by this link.
We have%
\begin{equation}
p_{\left( s\right) i}g_{\mathrm{M}}^{ik}p_{\left( s\right) k}=g_{\mathrm{M}%
ik}\dot{f}_{\left( s\right) }^{i}\left( \tau \right) \dot{f}_{\left(
s\right) }^{k}\left( \tau \right) b^{2}\left( d\tau \right) ^{2}=m_{\left(
s\right) }^{2}c^{2}  \label{a3.10}
\end{equation}%
\begin{equation}
m_{\left( s\right) }=\frac{bd\tau }{c}\sqrt{g_{\mathrm{M}rs}\dot{f}_{\left(
s\right) }^{r}\dot{f}_{\left( s\right) }^{s}}  \label{a3.11}
\end{equation}%
Then it follows from (\ref{a3.8}) and (\ref{a3.11}), that%
\begin{equation}
p_{\left( s\right) }^{i}=\frac{m_{\left( s\right) }c\dot{f}^{i}}{\sqrt{g_{%
\mathrm{M}rs}\dot{f}_{\left( s\right) }^{r}\dot{f}_{\left( s\right) }^{s}}}
\label{a3.11a}
\end{equation}%
According (\ref{a3.7}) one obtains for the pointlike particles%
\begin{equation}
T^{ik}\left( x\right) =\dsum\limits_{s}\frac{m_{\left( s\right) }c\dot{f}%
_{\left( s\right) }^{i}\left( \tau \right) \dot{f}_{\left( s\right)
}^{k}\left( \tau \right) }{g_{\mathrm{M}rs}\dot{f}^{r}\left( \tau \right) 
\dot{f}^{s}\left( \tau \right) }\dprod\limits_{\alpha =1}^{\alpha =3}\delta
_{\alpha }\left( x^{\alpha }-f_{\left( s\right) }^{\alpha }\left( \tau
\right) \right)   \label{a3.12}
\end{equation}%
where $\delta $-function is defined by the relations%
\begin{equation}
\int_{V}\dprod\limits_{\alpha =1}^{\alpha =3}F\left( \mathbf{x}^{\prime
}\right) \delta _{\alpha }\left( x^{\prime \alpha }-f^{\alpha }\left( \tau
\right) \right) \sqrt{-g_{\mathrm{sp}}}d\mathbf{x}^{\prime }=\left\{ 
\begin{array}{ccc}
F\left( \mathbf{f}\left( \tau \right) \right)  & \text{if} & \mathbf{x}%
^{\prime }\in V \\ 
0 & \text{if} & \mathbf{x}^{\prime }\notin V%
\end{array}%
\right.   \label{a3.14}
\end{equation}%
Here 
\begin{equation}
g_{\mathrm{sp}}=\det \left\vert \left\vert \left( g_{\mathrm{M}}\right)
_{\alpha \beta }\right\vert \right\vert ,\qquad \alpha ,\beta =1,2,3
\label{a3.15a}
\end{equation}

Integral (\ref{a3.2}) over 
\[
d^{4}x^{\prime }=d^{3}\mathbf{x}^{\prime }dt^{\prime }=d^{3}\mathbf{x}%
^{\prime }\frac{dt^{\prime }}{d\tau }d\tau =d^{3}\mathbf{x}^{\prime }\dot{f}%
^{0}\left( \tau \right) d\tau 
\]%
can be presented in the form%
\begin{eqnarray}
\delta g^{ik}\left( x\right) &=&-\kappa \int G_{\mathrm{ret}}\left(
x,x^{\prime }\right) T^{ik}\left( x^{\prime }\right) \sqrt{-g_{\mathrm{M}}}%
d^{4}x^{\prime }  \nonumber \\
&=&-\kappa \int \dsum\limits_{s}\frac{m_{\left( s\right) }\dot{f}_{\left(
s\right) }^{i}\left( \tau \right) \dot{f}_{\left( s\right) }^{k}\left( \tau
\right) }{2\pi g_{\mathrm{M}rs}\dot{f}^{r}\dot{f}^{s}}\dprod\limits_{\alpha
=1}^{\alpha =3}\delta _{\alpha }\left( x^{\prime \alpha }-f_{\left( s\right)
}^{\alpha }\left( \tau \right) \right) \sqrt{-g_{\mathrm{M}}}d^{3}\mathbf{x}%
^{\prime }  \nonumber \\
&&\times \delta \left( \sigma _{\mathrm{M}}(x,f_{\left( s\right) }\left(
\tau \right) \right) \dot{f}_{\left( s\right) }^{0}\left( \tau \right) d\tau
\label{b3.1}
\end{eqnarray}%
Integration over $\mathbf{x}$ gives%
\begin{equation}
\delta g^{ik}\left( x\right) =-\kappa \int \dsum\limits_{s}\frac{m_{\left(
s\right) }\dot{f}_{\left( s\right) }^{i}\left( \tau \right) \dot{f}_{\left(
s\right) }^{k}\left( \tau \right) }{2\pi g_{\mathrm{M}rs}\dot{f}^{r}\left(
\tau \right) \dot{f}^{s}\left( \tau \right) }\sqrt{\frac{g_{\mathrm{M}}}{g_{%
\mathrm{sp}}}}\delta \left( 2\sigma _{\mathrm{M}}(x,f_{\left( s\right)
}\left( \tau \right) \right) \dot{f}_{\left( s\right) }^{0}\left( \tau
\right) d\tau  \label{b3.2}
\end{equation}%
Integration over $d\tau $ gives%
\begin{equation}
\delta g_{ik}\left( x\right) =-\frac{\kappa }{4\pi }\dsum\limits_{s}\frac{%
m_{\left( s\right) }g_{\mathrm{M}ij}\dot{f}_{\left( s\right) }^{j}\left(
\tau _{s}\right) g_{\mathrm{M}kl}\dot{f}_{\left( s\right) }^{l}\left( \tau
_{s}\right) }{g_{\mathrm{M}rs}\dot{f}^{r}\left( \tau _{s}\right) \dot{f}%
^{s}\left( \tau _{s}\right) \left\vert \frac{d}{d\tau }\sigma _{\mathrm{M}%
}\left( x,f\left( \tau _{s}\right) \right) \right\vert }\sqrt{\frac{g_{%
\mathrm{M}}}{g_{\mathrm{sp}}}}\dot{f}_{\left( s\right) }^{0}\left( \tau
_{s}\right)  \label{b3.3}
\end{equation}%
where $\tau _{s}=\tau _{s}\left( t,\mathbf{x}\right) $ is a root of the
equation%
\begin{equation}
2\sigma _{\mathrm{M}}\left( x,f\left( \tau _{s}\right) \right) =g_{\mathrm{M}%
ik}\left( x^{i}-f_{\left( s\right) }^{i}\left( \tau _{s}\right) \right)
\left( x^{k}-f_{\left( s\right) }^{k}\left( \tau _{s}\right) \right) =0
\label{b3.4}
\end{equation}%
which can be written in the form%
\begin{equation}
\sigma \left( P,P_{l}^{\prime }\right) =0  \label{b3.5}
\end{equation}%
We have%
\begin{equation}
\frac{d}{d\tau }\sigma _{\mathrm{M}}\left( x,f\left( \tau _{s}\right)
\right) =-g_{\mathrm{M}ik}\left( x^{i}-f_{\left( s\right) }^{i}\left( \tau
_{s}\right) \right) \dot{f}^{k}\left( \tau _{s}\right) =-\frac{\left( 
\mathbf{PP}_{l}^{\prime }.\mathbf{P}_{l}^{\prime }\mathbf{P}_{l+1}^{\prime
}\right) }{d\tau }  \label{b3.6}
\end{equation}

Using relations (\ref{f1.8}), (\ref{f1.8a}) one can rewrite the relation (%
\ref{b3.3}) in the form%
\begin{equation}
\delta g_{ik}\left( x\right) =-\frac{\kappa }{4\pi }\dsum\limits_{s}\frac{%
m_{\left( s\right) }\left( \mathbf{P}_{l}^{\prime }\mathbf{P}_{l+1}^{\prime
}.\mathbf{PQ}_{i}\right) \left( \mathbf{P}_{l}^{\prime }\mathbf{P}%
_{l+1}^{\prime }.\mathbf{PQ}_{k}\right) }{\left\vert \left( \mathbf{PP}%
_{l}^{\prime }.\mathbf{P}_{l}^{\prime }\mathbf{P}_{l+1}^{\prime }\right)
\right\vert \left( \mathbf{P}_{l}^{\prime }\mathbf{P}_{l+1}^{\prime }.%
\mathbf{P}_{l}^{\prime }\mathbf{P}_{l+1}^{\prime }\right) }\left( \mathbf{P}%
_{l}^{\prime }\mathbf{P}_{l+1}^{\prime }.\mathbf{PQ}_{k}\right) g_{\mathrm{M}%
}^{0k}\sqrt{\frac{g_{\mathrm{M}}}{g_{\mathrm{sp}}}}  \label{b3.7}
\end{equation}%
In the case, when all basic vectors $\mathbf{PQ}_{k}$ are unite and
orthogonal, the determinants $g_{\mathrm{M}}$ and $g_{\mathrm{sp}}$ are
connected by the relation 
\begin{equation}
g_{\mathrm{M}}=\det \left\vert \left\vert g_{\mathrm{M}ik}\right\vert
\right\vert =g_{\mathrm{sp}}\left( g_{\mathrm{M}}\right) _{00},\qquad \left(
g_{\mathrm{M}}\right) _{00}=\left( \mathbf{PQ}_{0}.\mathbf{PQ}_{0}\right)
=\left\vert \mathbf{PQ}_{0}\right\vert ^{2}  \label{b3.8}
\end{equation}%
The last multipliers of (\ref{b3.7}) can be written in the form%
\begin{equation}
g_{\mathrm{M}}^{00}\sqrt{\frac{g_{\mathrm{M}}}{g_{\mathrm{sp}}}}=\left(
\left( g_{\mathrm{M}}\right) _{00}\right) ^{-1}\sqrt{\left( g_{\mathrm{M}%
}\right) _{00}}=\frac{1}{\left\vert \mathbf{PQ}_{0}\right\vert }
\label{b3.9}
\end{equation}%
The constant $\kappa $ is connected with the gravitational constant $G$ by
means of the relation $\kappa =8\pi G/c^{2}$. Using (\ref{b3.9}) and (\ref%
{f1.9}), the relation (\ref{b3.7}) can be rewritten in terms of scalar
products 
\[
\delta g_{ik}\left( P\right) =\delta \left( \left( \mathbf{PQ}_{i}.\mathbf{PQ%
}_{k}\right) \right) 
\]%
\begin{equation}
=-\frac{2G}{c^{2}}\dsum\limits_{s}m_{\left( s\right) }\frac{\theta \left(
\left( \mathbf{P}_{l}^{\prime }\mathbf{P.PQ}_{0}\right) \right) }{\left( 
\mathbf{P}_{l}^{\prime }\mathbf{P.P}_{l}^{\prime }\mathbf{P}_{l+1}^{\prime
}\right) }\frac{\left( \mathbf{P}_{l}^{\prime }\mathbf{P}_{l+1}^{\prime }.%
\mathbf{PQ}_{i}\right) \left( \mathbf{P}_{l}^{\prime }\mathbf{P}%
_{l+1}^{\prime }.\mathbf{PQ}_{k}\right) \left( \mathbf{P}_{l}^{\prime }%
\mathbf{P}_{l+1}^{\prime }.\mathbf{PQ}_{0}\right) }{\left( \mathbf{P}%
_{l}^{\prime }\mathbf{P}_{l+1}^{\prime }.\mathbf{P}_{l}^{\prime }\mathbf{P}%
_{l+1}^{\prime }\right) \left\vert \mathbf{PQ}_{0}\right\vert }
\label{a3.18}
\end{equation}%
\begin{equation}
\sigma \left( P,P_{l}^{\prime }\right) =0  \label{a3.19}
\end{equation}%
\label{0end}where vectors $\mathbf{PQ}_{i},$ $i=0,1,2,3$ are basic vectors
of the coordinate system at the point $P$. Vector $\mathbf{PQ}_{0}$ is
timelike. The points $P_{l}^{\prime }$ and $P_{l+1}^{\prime }$ are on the
world line $\mathcal{L}_{\left( s\right) }$ of $s$th particle. The points $%
P_{l}^{\prime }$ and $P_{l+1}^{\prime }$ are separated by infinitesimal
distance. All scalar products are taken in the space-time geometry of
Minkowski. Besides, one uses the fact, that the metric tensor $g_{ik}\left(
P\right) $ at the point $P$ can be presented in the form 
\begin{equation}
g_{ik}\left( P\right) =\left( \mathbf{PQ}_{i}.\mathbf{PQ}_{k}\right) ,\qquad
i,k=0,1,2,3  \label{a3.20}
\end{equation}%
and the scalar product is expressed via the world function by means of the
relation (\ref{a2.4}).

To determine world function $\sigma $ from relations (\ref{a3.18}), (\ref%
{a3.19}), let us use the relation 
\begin{equation}
\left( \mathbf{PS}_{1}.\mathbf{PS}_{2}\right) =\sigma \left( P,S_{2}\right)
+\sigma \left( S_{1},P\right) -\sigma \left( P,P\right) -\sigma \left(
S_{1},S_{2}\right)  \label{a3.21}
\end{equation}%
As far as $\sigma \left( P,P\right) =0$, it may rewritten in the form 
\begin{equation}
\sigma \left( S_{1},S_{2}\right) =\sigma \left( P,S_{2}\right) +\sigma
\left( S_{1},P\right) -\left( \mathbf{PS}_{1}.\mathbf{PS}_{2}\right)
\label{a3.22}
\end{equation}%
Using (\ref{a2.5}) the relation (\ref{a3.22}) may be rewritten in terms of
scalar products%
\begin{equation}
\sigma \left( S_{1},S_{2}\right) =\frac{1}{2}\left( \left( \mathbf{PS}_{1}.%
\mathbf{PS}_{1}\right) +\left( \mathbf{PS}_{2}.\mathbf{PS}_{2}\right)
-2\left( \mathbf{PS}_{1}.\mathbf{PS}_{2}\right) \right)  \label{a3.23}
\end{equation}

Replacing $Q_{i}\rightarrow S_{1},Q_{k}\rightarrow S_{2}$ in relation (\ref%
{a3.18}) and substituting in (\ref{a3.23}), one obtains after transformations%
\begin{eqnarray}
\delta \sigma \left( S_{1},S_{2}\right) &=&-\frac{G}{c^{2}}%
\dsum\limits_{s}m_{\left( s\right) }\frac{\theta \left( \left( \mathbf{P}%
_{l}^{\prime }\mathbf{P.PQ}_{0}\right) \right) \left( \mathbf{P}_{l}^{\prime
}\mathbf{P}_{l+1}^{\prime }.\mathbf{PQ}_{0}\right) }{\left( \mathbf{P}%
_{l}^{\prime }\mathbf{P.P}_{l}^{\prime }\mathbf{P}_{l+1}^{\prime }\right)
\left\vert \mathbf{PQ}_{0}\right\vert }  \nonumber \\
&&\times \frac{\left( \left( \mathbf{P}_{l}^{\prime }\mathbf{P}%
_{l+1}^{\prime }.\mathbf{PS}_{1}\right) -\left( \mathbf{P}_{l}^{\prime }%
\mathbf{P}_{l+1}^{\prime }.\mathbf{PS}_{2}\right) \right) ^{2}}{\left( 
\mathbf{P}_{l}^{\prime }\mathbf{P}_{l+1}^{\prime }.\mathbf{P}_{l}^{\prime }%
\mathbf{P}_{l+1}^{\prime }\right) }  \label{a3.24}
\end{eqnarray}

The relations (\ref{a3.24}), (\ref{a3.19}) are completely geometric
relations, written in terms of the world function $\sigma _{\mathrm{M}}$ of
the Minkowski geometry. According to the deformation principle the relations
(\ref{a3.24}), (\ref{a3.19}) are valid in any physical space-time geometry
(i.e. for any world function $\sigma $). It means, that, if the space-time
geometry without additional particles is described by the world function $%
\sigma _{0}$, then appearance of additional particles perturbs the
space-time geometry, and it becomes to be described by the world function $%
\sigma =\sigma _{0}+\delta \sigma $, where perturbation $\delta \sigma $ of
the world function is determined by the relations (\ref{a3.24}), (\ref{a3.19}%
). Scalar products in rhs of (\ref{a3.24}) should be calculated by means of
the world function $\sigma $, which is unknown at first. As a result
equations (\ref{a3.24}), (\ref{a3.19}) form equations for determination of
the world function $\sigma $.

In the case of continuous distribution of particles the summation in (\ref%
{a3.24}) is to be substituted by integration over Lagrangian coordinates $%
\mathbf{\xi }$\textbf{,} labelling the perturbing particles. One obtains 
\begin{eqnarray}
\delta \left( S_{1},S_{2}\right) &=&-\frac{G}{c^{2}}\int_{V}\rho \left( 
\mathbf{\xi }\right) d\mathbf{\xi }\frac{\theta \left( \left( \mathbf{P}%
_{l}^{\prime }\mathbf{P.PQ}_{0}\right) \right) \left( \mathbf{P}_{l}^{\prime
}\mathbf{P}_{l+1}^{\prime }.\mathbf{PQ}_{0}\right) }{\left( \mathbf{P}%
_{l}^{\prime }\mathbf{P.P}_{l}^{\prime }\mathbf{P}_{l+1}^{\prime }\right)
\left\vert \mathbf{PQ}_{0}\right\vert }  \nonumber \\
&&\times \frac{\left( \left( \mathbf{P}_{l}^{\prime }\mathbf{P}%
_{l+1}^{\prime }.\mathbf{PS}_{1}\right) -\left( \mathbf{P}_{l}^{\prime }%
\mathbf{P}_{l+1}^{\prime }.\mathbf{PS}_{2}\right) \right) ^{2}}{\left( 
\mathbf{P}_{l}^{\prime }\mathbf{P}_{l+1}^{\prime }.\mathbf{P}_{l}^{\prime }%
\mathbf{P}_{l+1}^{\prime }\right) }  \label{a3.25}
\end{eqnarray}%
where the total mass $M$ is defined by the relation 
\begin{equation}
\int_{V}\rho \left( \mathbf{\xi }\right) d\mathbf{\xi =}M  \label{a3.25a}
\end{equation}%
The points $S_{1}$ and $S_{2}$ are arbitrary points of the space-time. The
point $P$ is to be some function $P=P\left( S_{1},S_{2}\right) $ of points $%
S_{1}$ and $S_{2}$, which determine position of the point $P$ between the
points $S_{1}$ and $S_{2}$. As far as $\delta \sigma \left(
S_{1},S_{2}\right) $ is a symmetric function of arguments $S_{1},S_{2}$, the
function $P\left( S_{1},S_{2}\right) $ is to be symmetric also, as it
follows from (\ref{a3.25}) 
\begin{equation}
P\left( S_{1},S_{2}\right) =P\left( S_{2},S_{1}\right)  \label{a3.26}
\end{equation}%
The function $P\left( S_{1},S_{2}\right) $ is to be determined by completely
geometric relations, written in terms of the world function $\sigma $. In
the geometry of Minkowski these relations can have the form%
\begin{equation}
4\sigma \left( P,S_{1}\right) =\sigma \left( S_{1},S_{2}\right) ,\qquad
4\sigma \left( P,S_{2}\right) =\sigma \left( S_{1},S_{2}\right)
\label{a3.28}
\end{equation}%
Two equations (\ref{a3.28}) have an unique solution in the space-time of
Minkowski, if the distance $S_{1}S_{2}$ is timelike. The point $P$ lies in
the middle of the segment $S_{1}S_{2}$. Indeed, let the world function $%
\sigma \left( S_{1},S_{2}\right) >0$, and the vector $\mathbf{S}_{1}\mathbf{S%
}_{2}$ be timelike. Choosing coordinate system in such a way, that 
\begin{equation}
S_{1}=\left\{ 0,0,0,0\right\} \qquad S_{2}=\left\{ a_{0},0,0,0\right\}
,\qquad P=\left\{ t,x^{1},x^{2},x^{3}\right\} ,  \label{a3.29}
\end{equation}%
the equations (\ref{a3.28}) are written in the form%
\begin{equation}
c^{2}t^{2}-\mathbf{x}^{2}=c^{2}\frac{a_{0}^{2}}{4},\qquad c^{2}\left(
t-a_{0}\right) ^{2}-\left( x^{1}\right) ^{2}-\left( x^{2}\right) ^{2}-\left(
x^{3}\right) ^{2}=c^{2}\frac{a^{2}}{4}  \label{a3.40}
\end{equation}%
Taking difference of the two equations, one obtains%
\begin{equation}
c^{2}\left( 2a_{0}t-a_{0}^{2}\right) =0,\qquad t=\frac{a_{0}}{2}
\label{a3.41}
\end{equation}%
Substituting$\ $ $t=\frac{a_{0}}{2}$ in the second equation (\ref{a3.40}),
one obtains%
\begin{equation}
\mathbf{x}^{2}=0,\qquad P=\left\{ \frac{a_{0}}{2},0,0,0\right\}
\label{a3.42}
\end{equation}

If the segment $\mathbf{S}_{1}\mathbf{S}_{2}$ is spacelike, the coordinate
system may be chosen in such a way, that 
\begin{equation}
S_{1}=\left\{ 0,0,0,0\right\} \qquad S_{2}=\left\{ 0,a_{1},0,0,0\right\}
,\qquad P=\left\{ t,x^{1},x^{2},x^{3}\right\}  \label{a3.43}
\end{equation}%
In this case the solution of equations (\ref{a3.28}) looks as follows%
\begin{equation}
P=\left\{ \sqrt{a_{2}^{2}+a_{3}^{2}},a_{1},a_{2},a_{3}\right\}  \label{a3.44}
\end{equation}%
where $a_{2}$ and $a_{3}$ are arbitrary quantities. Ambiguity of the
solution is a manifestation of the Minkowski geometry multivariance with
respect to spacelike vectors.

In this connection let us note, that at description of the particle motion
only values of the world function for timelike intervals are important.
Maybe, ambiguity of the world function for spacelike interval is not so
important for description of the space-time properties. At any rate, one
uses timelike world lines for description of particle dynamics.

\section{World function of non-rotating body}

Let us consider a physical body, which is concentrated in a space volume $V$%
. Its density is $\rho \left( \mathbf{\xi }\right) $, where $\mathbf{\xi }$
are Lagrangian coordinates of the body points. The body does not rotate. We
shall use the inertial coordinate system $x=\left\{ t,\mathbf{x}\right\}
=\left\{ t,x^{1},x^{2},x^{3}\right\} $.

We shall search for solution of equations (\ref{a3.25}), (\ref{a3.19}) for
the body in the form of a second order polynomial of $\left(
t_{1}-t_{2}\right) $%
\begin{equation}
\sigma \left( t_{1},\mathbf{y}_{1};t_{2},\mathbf{y}_{2}\right) =\frac{1}{2}%
A\left( \mathbf{y}_{1},\mathbf{y}_{2}\right) c^{2}\left( t_{2}-t_{1}\right)
^{2}+B\left( \mathbf{y}_{1},\mathbf{y}_{2}\right) c\left( t_{2}-t_{1}\right)
+C\left( \mathbf{y}_{1},\mathbf{y}_{2}\right)  \label{a5.4}
\end{equation}%
\begin{equation}
A\left( \mathbf{y}_{1},\mathbf{y}_{2}\right) =1-V\left( \mathbf{y}_{1},%
\mathbf{y}_{2}\right) ,\qquad C\left( \mathbf{y}_{1},\mathbf{y}_{2}\right) =-%
\frac{1}{2}\left( \mathbf{y}_{1}-\mathbf{y}_{2}\right) ^{2}+\delta C\left( 
\mathbf{y}_{1},\mathbf{y}_{2}\right)  \label{a5.5}
\end{equation}%
where functions $A,B$ and $C$ should be determined as a result of solution
of equations (\ref{a3.25}), (\ref{a3.19}). In the zeroth order
approximation, when the space-time is the space of Minkowski, one has%
\begin{equation}
A_{0}\left( \mathbf{y}_{1},\mathbf{y}_{2}\right) =1,\quad V_{0}\left( 
\mathbf{y}_{1},\mathbf{y}_{2}\right) =0,\quad B_{0}\left( \mathbf{y}_{1},%
\mathbf{y}_{2}\right) =0,\quad \delta C_{0}\left( \mathbf{y}_{1},\mathbf{y}%
_{2}\right) =0  \label{a5.6}
\end{equation}

Let coordinates of points have the form%
\begin{eqnarray}
P_{l}^{\prime } &=&\left\{ t-\frac{r}{c},\mathbf{\xi }\right\} ,\qquad
P_{l+1}^{\prime }=\left\{ t-\frac{r}{c}+dT,\mathbf{\xi }\right\} ,  \nonumber
\\
P &=&\left\{ t,\mathbf{x}\right\} \qquad S_{1}=\left\{ t_{1},\mathbf{y}%
_{1}\right\} \qquad S_{2}=\left\{ t_{2},\mathbf{y}_{2}\right\}  \nonumber \\
Q_{0} &=&\left\{ t+dt,\mathbf{x}\right\} ,\qquad Q_{1}=\left\{
t,x^{1}+dx^{1},x^{2},x^{3}\right\} ,  \nonumber \\
Q_{2} &=&\left\{ t,x^{1},x^{2}+dx^{2},x^{3}\right\} ,\qquad Q_{3}=\left\{
t,x^{1},x^{2},x^{3}+dx^{3}\right\}  \label{a5.7}
\end{eqnarray}%
where coordinates $\mathbf{\xi }$ label points of the body.

Vectors $\mathbf{PQ}$ in scalar products of the expression (\ref{a3.25}) are
described by coordinates of points $P$ and \ $Q$: $\mathbf{PQ=}\left\{
x\left( P\right) ;x\left( Q\right) \right\} $, where $x\left( P\right) $ are
coordinates of the point $P$. By means of (\ref{a5.7}) we have the following
coordinates for vectors in (\ref{a3.25}): 
\begin{eqnarray}
\mathbf{P}_{l}^{\prime }\mathbf{P} &=&\left\{ t-\frac{r}{c},\mathbf{\xi };t,%
\mathbf{x}\right\} ,\quad \mathbf{PQ}_{0}=\left\{ t,\mathbf{x;}t+dt,\mathbf{x%
}\right\} ,\quad \mathbf{P}_{l}^{\prime }\mathbf{P}_{l+1}^{\prime }=\left\{
t-\frac{r}{c},\mathbf{\xi ;}t-\frac{r}{c}+dT,\mathbf{\xi }\right\} ,\quad 
\nonumber \\
\mathbf{PS}_{1} &=&\left\{ t,\mathbf{x;}t_{1},\mathbf{y}_{1}\right\} ,\quad 
\mathbf{PS}_{2}=\left\{ t,\mathbf{x;}t_{2},\mathbf{y}_{2}\right\}
\label{a5.8}
\end{eqnarray}%
The quantity $dT$ is supposed to be infinitesimal.

In the first approximation the world function has the form%
\begin{equation}
\sigma _{1}\left( t_{1},\mathbf{y}_{1};t_{2},\mathbf{y}_{2}\right) =\frac{1}{%
2}A_{1}\left( \mathbf{y}_{1},\mathbf{y}_{2}\right) c^{2}\left(
t_{2}-t_{1}\right) ^{2}+B_{1}\left( \mathbf{y}_{1},\mathbf{y}_{2}\right)
c\left( t_{2}-t_{1}\right) +C_{1}\left( \mathbf{y}_{1},\mathbf{y}_{2}\right)
\label{a5.9}
\end{equation}

As far as $\sigma _{1}=\sigma _{\mathrm{M}}+\delta \sigma _{1}$, it follows
from (\ref{a5.9}) 
\begin{equation}
\delta \sigma _{1}\left( t_{1},\mathbf{y}_{1};t_{2},\mathbf{y}_{2}\right) =-%
\frac{1}{2}V_{1}\left( \mathbf{y}_{1},\mathbf{y}_{2}\right) c^{2}\left(
t_{2}-t_{1}\right) ^{2}+B_{1}\left( \mathbf{y}_{1},\mathbf{y}_{2}\right)
c\left( t_{2}-t_{1}\right) +\delta C_{1}\left( \mathbf{y}_{1},\mathbf{y}%
_{2}\right)  \label{a5.9a}
\end{equation}%
where%
\begin{equation}
\delta C_{1}\left( \mathbf{y}_{1},\mathbf{y}_{2}\right) =C_{1}\left( \mathbf{%
y}_{1},\mathbf{y}_{2}\right) +\frac{1}{2}\left( \mathbf{y}_{2}-\mathbf{y}%
_{1}\right) ^{2}  \label{a5.9b}
\end{equation}%
At $t_{2}\rightarrow t_{1}$ the world function $\sigma _{1}\left( t_{1},%
\mathbf{y}_{1};t_{2},\mathbf{y}_{1}\right) $ tends to $0$. Taking into
account the symmetry of the world function with respect to transposition $%
\left( t_{1},\mathbf{y}_{1}\right) \leftrightarrow \left( t_{2},\mathbf{y}%
_{1}\right) $, we conclude, that%
\begin{equation}
C_{1}\left( \mathbf{y}_{1},\mathbf{y}_{1}\right) =0,\quad B_{1}\left( 
\mathbf{y}_{1},\mathbf{y}_{1}\right) =0  \label{a5.10}
\end{equation}

According to equation (\ref{a3.19}), we obtain from (\ref{a5.9}), (\ref{a5.8}%
) 
\begin{equation}
\frac{1}{2}\left\vert \mathbf{P}_{l}^{\prime }\mathbf{P}\right\vert ^{2}=%
\frac{1}{2}A_{1}\left( \mathbf{\xi },\mathbf{x}\right) c^{2}\left( \frac{r}{c%
}\right) ^{2}+B_{1}\left( \mathbf{\xi },\mathbf{x}\right) r+C_{1}\left( 
\mathbf{\xi ,x}\right) =0  \label{a5.11}
\end{equation}%
Solution of (\ref{a5.11}) has the form%
\begin{eqnarray}
r &=&\frac{-B_{1}\left( \mathbf{\xi },\mathbf{x}\right) +\sqrt{%
B_{1}^{2}\left( \mathbf{\xi },\mathbf{x}\right) -2C_{1}\left( \mathbf{\xi ,x}%
\right) A_{1}\left( \mathbf{\xi },\mathbf{x}\right) }}{A_{1}\left( \mathbf{%
\xi },\mathbf{x}\right) }  \nonumber \\
&=&-\frac{2C_{1}\left( \mathbf{\xi ,x}\right) }{B_{1}\left( \mathbf{\xi },%
\mathbf{x}\right) +\sqrt{B_{1}^{2}\left( \mathbf{\xi },\mathbf{x}\right)
-2C_{1}\left( \mathbf{\xi ,x}\right) A_{1}\left( \mathbf{\xi },\mathbf{x}%
\right) }}  \label{a5.12}
\end{eqnarray}

Calculation of other scalar products gives the results%
\begin{equation}
\left\vert \mathbf{PQ}_{0}\right\vert ^{2}=A_{1}\left( \mathbf{y}_{1},%
\mathbf{y}_{2}\right) c^{2}\left( dt\right) ^{2}  \label{a5.14}
\end{equation}%
\begin{equation}
\left( \mathbf{P}_{l}^{\prime }\mathbf{P}_{l+1}^{\prime }.\mathbf{P}%
_{l}^{\prime }\mathbf{P}_{l+1}^{\prime }\right) =A_{1}\left( \mathbf{\xi
,\xi }\right) c^{2}\left( dT\right) ^{2}  \label{a5.15}
\end{equation}%
\begin{eqnarray}
\left( \mathbf{P}_{l}^{\prime }\mathbf{P}_{l+1}^{\prime }.\mathbf{P}%
_{l}^{\prime }\mathbf{P}\right) &=&\sigma \left( P_{l}^{\prime },P\right)
+\sigma \left( P_{l+1}^{\prime },P_{l}^{\prime }\right) -0-\sigma \left(
P_{l+1}^{\prime },P\right)  \nonumber \\
&=&\sigma \left( P_{l+1}^{\prime },P_{l}^{\prime }\right) -\sigma \left(
P_{l+1}^{\prime },P\right)  \label{a5.16}
\end{eqnarray}%
\begin{eqnarray}
\sigma \left( P_{l+1}^{\prime },P\right) &=&\sigma \left( t-\frac{r}{c}+dT,%
\mathbf{\xi ;}t,\mathbf{x}\right)  \label{a5.17} \\
&=&\frac{1}{2}A_{1}\left( \mathbf{\xi },\mathbf{x}\right) c^{2}\left( \frac{r%
}{c}-dT\right) ^{2}+cB_{1}\left( \mathbf{\xi },\mathbf{x}\right) \left( 
\frac{r}{c}-dT\right) +C_{1}\left( \mathbf{\xi },\mathbf{x}\right)  \nonumber
\end{eqnarray}%
\begin{eqnarray}
&&\left( \mathbf{P}_{l}^{\prime }\mathbf{P}_{l+1}^{\prime }.\mathbf{P}%
_{l}^{\prime }\mathbf{P}\right) =\sigma \left( P_{l+1}^{\prime
},P_{l}^{\prime }\right) -\sigma \left( P_{l+1}^{\prime },P\right)
\label{a5.18} \\
&=&\frac{1}{2}A_{1}\left( \mathbf{\xi ,\xi }\right) c^{2}\left( dT\right)
^{2}-\left( \frac{1}{2}A_{1}\left( \mathbf{\xi },\mathbf{x}\right)
c^{2}\left( \frac{r}{c}-dT\right) ^{2}+cB_{1}\left( \mathbf{\xi },\mathbf{x}%
\right) \left( \frac{r}{c}-dT\right) +C_{1}\left( \mathbf{\xi },\mathbf{x}%
\right) \right)  \nonumber \\
&=&-\frac{1}{2}A_{1}\left( \mathbf{\xi },\mathbf{x}\right) r^{2}+A_{1}\left( 
\mathbf{\xi },\mathbf{x}\right) crdT-B_{1}\left( \mathbf{\xi },\mathbf{x}%
\right) r+B_{1}\left( \mathbf{\xi },\mathbf{x}\right) cdT-C_{1}\left( 
\mathbf{\xi },\mathbf{x}\right) +\mathcal{O}\left( dT^{2}\right)  \nonumber
\end{eqnarray}%
Taking into account, the relation (\ref{a5.11}) one obtains%
\begin{equation}
\left( \mathbf{P}_{l}^{\prime }\mathbf{P}_{l+1}^{\prime }.\mathbf{P}%
_{l}^{\prime }\mathbf{P}\right) =_{1}\left( \mathbf{\xi },\mathbf{x}\right)
crdT+B_{1}\left( \mathbf{\xi },\mathbf{x}\right) cdT+\mathcal{O}\left(
dT^{2}\right)  \label{a5.19}
\end{equation}

Calculation of $\left( \mathbf{P}_{l}^{\prime }\mathbf{P}_{l+1}^{\prime }.%
\mathbf{PS}_{1}\right) -\left( \mathbf{P}_{l}^{\prime }\mathbf{P}%
_{l+1}^{\prime }.\mathbf{PS}_{2}\right) $ gives%
\begin{eqnarray}
&&\left( \mathbf{P}_{l}^{\prime }\mathbf{P}_{l+1}^{\prime }.\mathbf{PS}%
_{1}\right) -\left( \mathbf{P}_{l}^{\prime }\mathbf{P}_{l+1}^{\prime }.%
\mathbf{PS}_{2}\right)  \nonumber \\
&=&\sigma _{1}\left( P_{l}^{\prime },S_{1}\right) +\sigma _{1}\left(
P_{l+1}^{\prime },P\right) -\sigma _{1}\left( P_{l}^{\prime },P\right)
-\sigma _{1}\left( P_{l+1}^{\prime },S_{1}\right)  \nonumber \\
&&-\left( \sigma _{1}\left( P_{l}^{\prime },S_{2}\right) +\sigma _{1}\left(
P_{l+1}^{\prime },P\right) -\sigma _{1}\left( P_{l}^{\prime },P\right)
-\sigma _{1}\left( P_{l+1}^{\prime },S_{2}\right) \right)  \nonumber \\
&=&\sigma _{1}\left( P_{l}^{\prime },S_{1}\right) -\sigma _{1}\left(
P_{l+1}^{\prime },S_{1}\right) -\left( \sigma _{1}\left( P_{l}^{\prime
},S_{2}\right) -\sigma _{1}\left( P_{l+1}^{\prime },S_{2}\right) \right) 
\nonumber \\
&=&-dT\frac{\partial }{\partial dT}\sigma _{1}\left( P_{l+1}^{\prime
},S_{1}\right) +dT\frac{\partial }{\partial dT}\sigma _{1}\left(
P_{l+1}^{\prime },S_{2}\right)  \nonumber \\
&=&dT\frac{\partial }{\partial dT}\left( \sigma _{1}\left( P_{l+1}^{\prime
},S_{2}\right) -\sigma _{1}\left( P_{l+1}^{\prime },S_{1}\right) \right) +%
\mathcal{O}\left( dT^{2}\right)  \label{a5.20}
\end{eqnarray}%
Using (\ref{a5.4}) and 
\begin{equation}
\mathbf{P}_{l}^{\prime }\mathbf{S}_{1}=\left\{ t-\frac{r}{c},\mathbf{\xi ;}%
t_{1},\mathbf{y}_{1}\right\} ,\qquad \mathbf{P}_{l+1}^{\prime }\mathbf{S}%
_{1}=\left\{ t-\frac{r}{c}+dT,\mathbf{\xi ;}t_{1},\mathbf{y}_{1}\right\}
\label{a5.21}
\end{equation}%
one obtains from (\ref{a5.20})%
\begin{eqnarray}
&&\left( \mathbf{P}_{l}^{\prime }\mathbf{P}_{l+1}^{\prime }.\mathbf{PS}%
_{1}\right) -\left( \mathbf{P}_{l}^{\prime }\mathbf{P}_{l+1}^{\prime }.%
\mathbf{PS}_{2}\right)  \nonumber \\
&=&+A_{1}\left( \mathbf{\xi },\mathbf{y}_{1}\right) c^{2}t_{1}dT-A_{1}\left( 
\mathbf{\xi },\mathbf{y}_{2}\right) c^{2}t_{2}dT+\left( A_{1}\left( \mathbf{%
\xi },\mathbf{y}_{2}\right) -A_{1}\left( \mathbf{\xi },\mathbf{y}_{1}\right)
\right) c^{2}\left( t-\frac{r}{c}\right) dT  \nonumber \\
&&+\left( B_{1}\left( \mathbf{\xi },\mathbf{y}_{1}\right) -B_{1}\left( 
\mathbf{\xi },\mathbf{y}_{2}\right) \right) cdT+\mathcal{O}\left(
dT^{2}\right)  \label{a5.22}
\end{eqnarray}

Let us take into account, that the time coordinate $t$ of the point $P$ has
the form%
\begin{equation}
t=\frac{t_{1}+t_{2}}{2}  \label{a5.23}
\end{equation}%
The relation (\ref{a5.22}) takes the form%
\begin{eqnarray}
&&\left( \mathbf{P}_{l}^{\prime }\mathbf{P}_{l+1}^{\prime }.\mathbf{PS}%
_{1}\right) -\left( \mathbf{P}_{l}^{\prime }\mathbf{P}_{l+1}^{\prime }.%
\mathbf{PS}_{2}\right)  \nonumber \\
&=&+\frac{1}{2}\left( A_{1}\left( \mathbf{\xi },\mathbf{y}_{1}\right)
+A_{1}\left( \mathbf{\xi },\mathbf{y}_{2}\right) \right) c^{2}\left(
t_{1}-t_{2}\right) dT-\left( A_{1}\left( \mathbf{\xi },\mathbf{y}_{2}\right)
-A_{1}\left( \mathbf{\xi },\mathbf{y}_{1}\right) \right) rcdT  \nonumber \\
&&+\left( B_{1}\left( \mathbf{\xi },\mathbf{y}_{1}\right) -B_{1}\left( 
\mathbf{\xi },\mathbf{y}_{2}\right) \right) cdT  \label{a5.24}
\end{eqnarray}%
Using (\ref{a5.5}), the relation (\ref{a5.24}) can be written in the form %
\label{0PS}%
\begin{eqnarray}
&&\left( \mathbf{P}_{l}^{\prime }\mathbf{P}_{l+1}^{\prime }.\mathbf{PS}%
_{1}\right) -\left( \mathbf{P}_{l}^{\prime }\mathbf{P}_{l+1}^{\prime }.%
\mathbf{PS}_{2}\right)  \nonumber \\
&=&\left( 1-\frac{1}{2}\left( V_{1}\left( \mathbf{\xi },\mathbf{y}%
_{2}\right) +V_{1}\left( \mathbf{\xi },\mathbf{y}_{1}\right) \right) \right)
c^{2}\left( t_{1}-t_{2}\right) dT+\left( V_{1}\left( \mathbf{\xi },\mathbf{y}%
_{2}\right) -V_{1}\left( \mathbf{\xi },\mathbf{y}_{1}\right) \right) rcdT 
\nonumber \\
&&+\left( B_{1}\left( \mathbf{\xi },\mathbf{y}_{1}\right) -B_{1}\left( 
\mathbf{\xi },\mathbf{y}_{2}\right) \right) cdT  \label{a5.25}
\end{eqnarray}

Calculation gives the following result for scalar product $\left( \mathbf{P}%
_{l}^{\prime }\mathbf{P.PQ}_{0}\right) $ 
\begin{equation}
\left( \mathbf{P}_{l}^{\prime }\mathbf{P.PQ}_{0}\right) =\left( 2A_{1}\left( 
\mathbf{\xi ,x}\right) r+B_{1}\left( \mathbf{\xi ,x}\right) \right) cdt
\label{a5.26}
\end{equation}%
This scalar product is positive, if $r$, defined by the relation (\ref{a5.12}%
), is positive and $A_{1}\left( \mathbf{\xi ,x}\right) >0$.

After substitution of expressions (\ref{a5.10}), (\ref{a5.11}), (\ref{a5.15}%
), (\ref{a5.19}) and (\ref{a5.25}), the expression (\ref{a3.25}) takes the
form 
\begin{eqnarray}
&&\delta \sigma \left( S_{1},S_{2}\right)  \nonumber \\
&=&-\frac{G}{c^{2}}\int_{V}\rho \left( \mathbf{\xi }\right) d\mathbf{\xi }%
\frac{\theta \left( \left( \mathbf{P}_{l}^{\prime }\mathbf{P.PQ}_{0}\right)
\right) A_{1}\left( \mathbf{\xi ,x}\right) c^{2}dtdT}{\sqrt{A_{1}\left( 
\mathbf{x},\mathbf{x}\right) }cdt}  \nonumber \\
&&\times \frac{\left( 
\begin{array}{c}
\left( 1-\frac{1}{2}\left( V_{1}\left( \mathbf{\xi },\mathbf{y}_{2}\right)
+V_{1}\left( \mathbf{\xi },\mathbf{y}_{1}\right) \right) \right) c\left(
t_{1}-t_{2}\right) \\ 
+\left( V_{1}\left( \mathbf{\xi },\mathbf{y}_{2}\right) -V_{1}\left( \mathbf{%
\xi },\mathbf{y}_{1}\right) \right) r+\left( B_{1}\left( \mathbf{\xi },%
\mathbf{y}_{1}\right) -B_{1}\left( \mathbf{\xi },\mathbf{y}_{2}\right)
\right)%
\end{array}%
\right) ^{2}\left( cdT\right) ^{2}}{\left( A_{1}\left( \mathbf{\xi },\mathbf{%
x}\right) r+B_{1}\left( \mathbf{\xi },\mathbf{x}\right) \right) A_{1}\left( 
\mathbf{\xi ,\xi }\right) c^{2}\left( dT\right) ^{2}cdT}  \label{a5.27}
\end{eqnarray}%
After cancellation of multiplier $dT$ and $dt$, we obtain%
\begin{eqnarray}
&&\delta \sigma \left( S_{1},S_{2}\right)  \nonumber \\
&=&-\frac{G}{c^{2}}\int_{V}\rho \left( \mathbf{\xi }\right) d\mathbf{\xi }%
\frac{A_{1}\left( \mathbf{\xi ,x}\right) }{\sqrt{A_{1}\left( \mathbf{x},%
\mathbf{x}\right) }\left( A_{1}\left( \mathbf{\xi },\mathbf{x}\right)
r+B_{1}\left( \mathbf{\xi },\mathbf{x}\right) \right) A_{1}\left( \mathbf{%
\xi ,\xi }\right) }  \nonumber \\
&&\times \left( 
\begin{array}{c}
\left( 1-\frac{1}{2}\left( V_{1}\left( \mathbf{\xi },\mathbf{y}_{2}\right)
+V_{1}\left( \mathbf{\xi },\mathbf{y}_{1}\right) \right) \right) c\left(
t_{1}-t_{2}\right) \\ 
+\left( V_{1}\left( \mathbf{\xi },\mathbf{y}_{2}\right) -V_{1}\left( \mathbf{%
\xi },\mathbf{y}_{1}\right) \right) r+\left( B_{1}\left( \mathbf{\xi },%
\mathbf{y}_{1}\right) -B_{1}\left( \mathbf{\xi },\mathbf{y}_{2}\right)
\right)%
\end{array}%
\right) ^{2}  \label{a5.28}
\end{eqnarray}%
where 
\begin{equation}
r=\frac{-B_{1}\left( \mathbf{\xi },\mathbf{x}\right) +\sqrt{B_{1}^{2}\left( 
\mathbf{\xi },\mathbf{x}\right) -2C_{1}\left( \mathbf{\xi ,x}\right)
A_{1}\left( \mathbf{\xi },\mathbf{x}\right) }}{A_{1}\left( \mathbf{\xi },%
\mathbf{x}\right) }  \label{a5.29}
\end{equation}

One can see, that rhs of (\ref{a5.28}) is the second order polynomial of $%
\left( t_{1}-t_{2}\right) $. Thus, our supposition that the world function
is the second order polynomial of $\left( t_{1}-t_{2}\right) $ is not
changed after variation of the world function under influence of additional
particles. 
\begin{equation}
\delta \sigma _{2}\left( t_{1},\mathbf{y}_{1};t_{2},\mathbf{y}_{2}\right) =-%
\frac{1}{2}V_{2}\left( \mathbf{y}_{1},\mathbf{y}_{2}\right) c^{2}\left(
t_{2}-t_{1}\right) ^{2}+B_{2}\left( \mathbf{y}_{1},\mathbf{y}_{2}\right)
c\left( t_{2}-t_{1}\right) +\delta C_{2}\left( \mathbf{y}_{1},\mathbf{y}%
_{2}\right)  \label{a5.30}
\end{equation}%
On the other side, it follows from (\ref{a5.28})%
\begin{eqnarray}
&&\delta \sigma _{2}\left( S_{1},S_{2}\right)  \nonumber \\
&=&-\int_{V}D\left( \mathbf{x,\xi }\right) \left( 1-\frac{1}{2}\left(
V_{1}\left( \mathbf{\xi },\mathbf{y}_{2}\right) +V_{1}\left( \mathbf{\xi },%
\mathbf{y}_{1}\right) \right) \right) ^{2}c^{2}\left( t_{1}-t_{2}\right)
^{2}d\mathbf{\xi }  \nonumber \\
&&-2\int_{V}D\left( \mathbf{x,\xi }\right) \left( 1-\frac{1}{2}\left(
V_{1}\left( \mathbf{\xi },\mathbf{y}_{2}\right) +V_{1}\left( \mathbf{\xi },%
\mathbf{y}_{1}\right) \right) \right) c\left( t_{1}-t_{2}\right)  \nonumber
\\
&&\times \left( \left( V_{1}\left( \mathbf{\xi },\mathbf{y}_{2}\right)
-V_{1}\left( \mathbf{\xi },\mathbf{y}_{1}\right) \right) r+\left(
B_{1}\left( \mathbf{\xi },\mathbf{y}_{1}\right) -B_{1}\left( \mathbf{\xi },%
\mathbf{y}_{2}\right) \right) \right) d\mathbf{\xi }  \nonumber \\
&&-\int_{V}D\left( \mathbf{x,\xi }\right) \left( V_{1}\left( \mathbf{\xi },%
\mathbf{y}_{2}\right) -V_{1}\left( \mathbf{\xi },\mathbf{y}_{1}\right)
\right) r+\left( B_{1}\left( \mathbf{\xi },\mathbf{y}_{1}\right)
-B_{1}\left( \mathbf{\xi },\mathbf{y}_{2}\right) \right) ^{2}d\mathbf{\xi }
\label{a5.30a}
\end{eqnarray}%
where 
\begin{eqnarray}
D\left( \mathbf{x,\xi }\right) &=&\frac{G}{c^{2}}\frac{\rho \left( \mathbf{%
\xi }\right) A_{1}\left( \mathbf{\xi ,x}\right) }{A_{1}\left( \mathbf{\xi
,\xi }\right) \sqrt{A_{1}\left( \mathbf{x},\mathbf{x}\right) }\left(
A_{1}\left( \mathbf{\xi },\mathbf{x}\right) r+B_{1}\left( \mathbf{\xi },%
\mathbf{x}\right) \right) }  \nonumber \\
&=&\frac{G}{c^{2}}\frac{\rho \left( \mathbf{\xi }\right) A_{1}\left( \mathbf{%
\xi ,x}\right) }{A_{1}\left( \mathbf{\xi ,\xi }\right) \sqrt{A_{1}\left( 
\mathbf{x},\mathbf{x}\right) }\sqrt{B_{1}^{2}\left( \mathbf{\xi },\mathbf{x}%
\right) -2C_{1}\left( \mathbf{\xi ,x}\right) A_{1}\left( \mathbf{\xi },%
\mathbf{x}\right) }}  \label{a5.33}
\end{eqnarray}%
Here%
\begin{equation}
C_{1}\left( \mathbf{\xi ,x}\right) =-\frac{1}{2}\left( \mathbf{x-\xi }%
\right) ^{2}+\delta C_{1}\left( \mathbf{\xi ,x}\right)  \label{a5.33a}
\end{equation}%
Comparing (\ref{a5.30}) and (\ref{a5.30a}), one concludes%
\begin{equation}
V_{2}\left( \mathbf{y}_{1},\mathbf{y}_{2}\right) =2\int_{V}D\left( \mathbf{%
x,\xi }\right) \left( 1-\frac{1}{2}\left( V_{1}\left( \mathbf{\xi },\mathbf{y%
}_{2}\right) +V_{1}\left( \mathbf{\xi },\mathbf{y}_{1}\right) \right)
\right) ^{2}d\mathbf{\xi }  \label{a5.31}
\end{equation}%
\[
B_{2}\left( \mathbf{y}_{1},\mathbf{y}_{2}\right) =-2\int_{V}D\left( \mathbf{%
x,\xi }\right) \left( 1-\frac{1}{2}\left( V_{1}\left( \mathbf{\xi },\mathbf{y%
}_{2}\right) +V_{1}\left( \mathbf{\xi },\mathbf{y}_{1}\right) \right)
\right) 
\]%
\begin{equation}
\times \left( \left( V_{1}\left( \mathbf{\xi },\mathbf{y}_{2}\right)
-V_{1}\left( \mathbf{\xi },\mathbf{y}_{1}\right) \right) r+\left(
B_{1}\left( \mathbf{\xi },\mathbf{y}_{1}\right) -B_{1}\left( \mathbf{\xi },%
\mathbf{y}_{2}\right) \right) \right) d\mathbf{\xi }  \label{a5.34}
\end{equation}%
\begin{equation}
\delta C_{2}\left( \mathbf{y}_{1},\mathbf{y}_{2}\right) =-\int_{V}D\left( 
\mathbf{x,\xi }\right) \left( V_{1}\left( \mathbf{\xi },\mathbf{y}%
_{2}\right) -V_{1}\left( \mathbf{\xi },\mathbf{y}_{1}\right) \right)
r+\left( B_{1}\left( \mathbf{\xi },\mathbf{y}_{1}\right) -B_{1}\left( 
\mathbf{\xi },\mathbf{y}_{2}\right) \right) ^{2}d\mathbf{\xi }
\label{a5.35a}
\end{equation}

Substituting $V_{2},B_{2},\delta C_{2}\ $in rhs of equations (\ref{a5.31}) -
(\ref{a5.35a}) instead of $V_{1},B_{1},\delta C_{1}$, we obtain the
quantities $V_{3},B_{3},\delta C_{3}$. Continuing this process, we obtain in
the limit, that the quantities $V_{n},B_{n},\delta C_{n}$, appear to be
equal in both sides of equations (\ref{a5.31}) - (\ref{a5.35a}). In the
developed form these equations are written as follows 
\begin{equation}
V\left( \mathbf{y}_{1},\mathbf{y}_{2}\right) =\frac{2G}{c^{2}}\int_{V}\frac{%
\rho \left( \mathbf{\xi }\right) A\left( \mathbf{\xi ,x}\right) \left( 1-%
\frac{1}{2}\left( V\left( \mathbf{\xi },\mathbf{y}_{2}\right) +V\left( 
\mathbf{\xi },\mathbf{y}_{1}\right) \right) \right) ^{2}}{A\left( \mathbf{%
\xi ,\xi }\right) \sqrt{A\left( \mathbf{x},\mathbf{x}\right) }\sqrt{%
B^{2}\left( \mathbf{\xi },\mathbf{x}\right) +A\left( \mathbf{\xi },\mathbf{x}%
\right) \left( \left( \mathbf{x}-\mathbf{\xi }\right) ^{2}-2\delta C\left( 
\mathbf{\xi ,x}\right) \right) }}d\mathbf{\xi }  \label{a5.36}
\end{equation}%
\[
B\left( \mathbf{y}_{1},\mathbf{y}_{2}\right) =-2\frac{G}{c^{2}}\int_{V}\frac{%
\rho \left( \mathbf{\xi }\right) A\left( \mathbf{\xi ,x}\right) \left( 1-%
\frac{1}{2}\left( V\left( \mathbf{\xi },\mathbf{y}_{2}\right) +V\left( 
\mathbf{\xi },\mathbf{y}_{1}\right) \right) \right) }{A_{1}\left( \mathbf{%
\xi ,\xi }\right) \sqrt{A\left( \mathbf{x},\mathbf{x}\right) }\sqrt{%
B^{2}\left( \mathbf{\xi },\mathbf{x}\right) +A\left( \mathbf{\xi },\mathbf{x}%
\right) \left( \left( \mathbf{x}-\mathbf{\xi }\right) ^{2}-2\delta C\left( 
\mathbf{\xi ,x}\right) \right) }}d\mathbf{\xi } 
\]%
\begin{equation}
\times \left( \left( V\left( \mathbf{\xi },\mathbf{y}_{2}\right) -V\left( 
\mathbf{\xi },\mathbf{y}_{1}\right) \right) r+\left( B\left( \mathbf{\xi },%
\mathbf{y}_{1}\right) -B\left( \mathbf{\xi },\mathbf{y}_{2}\right) \right)
\right)  \label{a5.37}
\end{equation}%
\begin{equation}
\delta C\left( \mathbf{y}_{1},\mathbf{y}_{2}\right) =-\frac{G}{c^{2}}\int_{V}%
\frac{\rho \left( \mathbf{\xi }\right) A\left( \mathbf{\xi ,x}\right) \left(
\left( V\left( \mathbf{\xi },\mathbf{y}_{2}\right) -V\left( \mathbf{\xi },%
\mathbf{y}_{1}\right) \right) r+\left( B\left( \mathbf{\xi },\mathbf{y}%
_{1}\right) -B\left( \mathbf{\xi },\mathbf{y}_{2}\right) \right) \right) ^{2}%
}{A\left( \mathbf{\xi ,\xi }\right) \sqrt{A\left( \mathbf{x},\mathbf{x}%
\right) }\sqrt{B^{2}\left( \mathbf{\xi },\mathbf{x}\right) +A\left( \mathbf{%
\xi },\mathbf{x}\right) \left( \left( \mathbf{x}-\mathbf{\xi }\right)
^{2}-2\delta C\left( \mathbf{\xi ,x}\right) \right) }}d\mathbf{\xi }
\label{a5.38}
\end{equation}%
where%
\begin{equation}
\mathbf{x=}\frac{\mathbf{y}_{1}+\mathbf{y}_{2}}{2},\quad A\left( \mathbf{y}%
_{1},\mathbf{y}_{2}\right) =1-V\left( \mathbf{y}_{1},\mathbf{y}_{2}\right)
\label{a5.39}
\end{equation}%
\begin{equation}
r=\frac{-B\left( \mathbf{\xi },\mathbf{x}\right) +\sqrt{B^{2}\left( \mathbf{%
\xi },\mathbf{x}\right) +A\left( \mathbf{\xi },\mathbf{x}\right) \left(
\left( \mathbf{x-\xi }\right) ^{2}-2\delta C\left( \mathbf{\xi ,x}\right)
\right) }}{A\left( \mathbf{\xi },\mathbf{x}\right) }  \label{a5.40}
\end{equation}%
It follows from (\ref{a5.37}) - (\ref{a5.38}), that for $\mathbf{y}_{1}=%
\mathbf{y}_{2}=\mathbf{x}$%
\begin{equation}
B\left( \mathbf{x},\mathbf{x}\right) =0,\qquad \delta C\left( \mathbf{x},%
\mathbf{x}\right) =0  \label{a5.40a}
\end{equation}

Equations (\ref{a5.36}) - (\ref{a5.38}) are three integral equations for
determination of three quantities $V\left( \mathbf{y}_{1},\mathbf{y}%
_{2}\right) ,B\left( \mathbf{y}_{1},\mathbf{y}_{2}\right) ,\delta C\left( 
\mathbf{y}_{1},\mathbf{y}_{2}\right) $, which determine the world function%
\begin{eqnarray}
\sigma \left( t_{1},\mathbf{y}_{1};t_{2},\mathbf{y}_{2}\right) &=&\frac{1}{2}%
c^{2}\left( t_{2}-t_{1}\right) ^{2}-\frac{1}{2}\left( \mathbf{y}_{1}-\mathbf{%
y}_{2}\right) ^{2}-\frac{1}{2}V\left( \mathbf{y}_{1},\mathbf{y}_{2}\right)
c^{2}\left( t_{2}-t_{1}\right) ^{2}  \nonumber \\
&&+B\left( \mathbf{y}_{1},\mathbf{y}_{2}\right) c\left( t_{2}-t_{1}\right)
+\delta C_{1}\left( \mathbf{y}_{1},\mathbf{y}_{2}\right)  \label{a5.41}
\end{eqnarray}

\section{Dynamic equations for world \newline
function, generated by non-rotating sphere.}

Let the shape of the physical body be a sphere of radius $R$. Let us
introduce parameter $\varepsilon =r_{g}/R$, where $r_{g}=2GM/c^{2}$ is so
called gravitational radius. Let 
\begin{equation}
\varepsilon =\frac{2G}{c^{2}}\int_{V}\frac{\rho \left( \mathbf{\xi }\right) 
}{R}d\mathbf{\xi \ll }1  \label{a6.1}
\end{equation}%
Then it follows from equations (\ref{a5.38}) - (\ref{a5.40}), that%
\begin{equation}
V\left( \mathbf{y}_{1},\mathbf{y}_{2}\right) =\mathcal{O}\left( \varepsilon
\right) ,\qquad B\left( \mathbf{y}_{1},\mathbf{y}_{2}\right) =\mathcal{O}%
\left( \varepsilon ^{2}\right) ,\qquad \delta C\left( \mathbf{y}_{1},\mathbf{%
y}_{2}\right) =\mathcal{O}\left( \varepsilon ^{3}\right)  \label{a6.2}
\end{equation}

If $\varepsilon \ll 1$, equations (\ref{a5.36}) - (\ref{a5.38}) can be
solved by means of successive approximations. In the first approximation one
obtains 
\begin{equation}
V_{1}\left( \mathbf{y}_{1},\mathbf{y}_{2}\right) =\frac{2G}{c^{2}}\int_{V}%
\frac{\rho \left( \mathbf{\xi }\right) }{\sqrt{\left( \frac{\left\vert 
\mathbf{y}_{1}+\mathbf{y}_{2}\right\vert ^{2}}{4}-\mathbf{\xi }\right) ^{2}}}%
d\mathbf{\xi +}\mathcal{O}\left( \varepsilon ^{2}\right)  \label{a6.3}
\end{equation}%
\begin{equation}
B_{1}\left( \mathbf{y}_{1},\mathbf{y}_{2}\right) =0,\quad \delta C_{1}\left( 
\mathbf{y}_{1},\mathbf{y}_{2}\right) =0  \label{a6.4}
\end{equation}%
If $\rho \left( \mathbf{\xi }\right) $%
\begin{equation}
\rho \left( \mathbf{\xi }\right) =\left\{ 
\begin{array}{ccc}
\rho _{0} & \text{if} & \left\vert \mathbf{\xi }\right\vert <R \\ 
0 & \text{if} & \left\vert \mathbf{\xi }\right\vert >R%
\end{array}%
\right. ,\qquad \rho _{0}=\frac{3M}{4\pi R^{3}}=\text{const}  \label{a6.5}
\end{equation}%
where $M$ is the sphere mass, then%
\begin{equation}
V_{1}\left( \mathbf{y}_{1},\mathbf{y}_{2}\right) =\left\{ 
\begin{array}{ccc}
\frac{2GM}{c^{2}\left\vert \mathbf{x}\right\vert } & \text{if} & \left\vert 
\mathbf{x}\right\vert >R \\ 
3\frac{GM}{c^{2}R}-\frac{GM}{c^{2}R^{3}}\left\vert \mathbf{x}\right\vert ^{2}
& \text{if} & \left\vert \mathbf{x}\right\vert <R%
\end{array}%
\right. ,\qquad \mathbf{x}=\frac{\mathbf{y}_{1}+\mathbf{y}_{2}}{2}
\label{a6.6}
\end{equation}

In the second approximation one obtains%
\begin{equation}
V_{2}\left( \mathbf{y}_{1},\mathbf{y}_{2}\right) =\frac{2G}{c^{2}}\int_{V}%
\frac{\rho _{0}\left( \mathbf{\xi }\right) \sqrt{A_{1}\left( \mathbf{\xi ,x}%
\right) }\left( 1-\frac{1}{2}\left( V_{1}\left( \mathbf{\xi },\mathbf{y}%
_{2}\right) +V_{1}\left( \mathbf{\xi },\mathbf{y}_{1}\right) \right) \right)
^{2}}{A_{1}\left( \mathbf{\xi ,\xi }\right) \sqrt{A_{1}\left( \mathbf{x},%
\mathbf{x}\right) }\sqrt{\left( \mathbf{x}-\mathbf{\xi }\right) ^{2}}}d%
\mathbf{\xi +}\mathcal{O}\left( \varepsilon ^{3}\right)  \label{a6.7}
\end{equation}%
\begin{eqnarray}
B_{2}\left( \mathbf{y}_{1},\mathbf{y}_{2}\right) &=&-2\frac{G}{c^{2}}\int_{V}%
\frac{\rho _{0}\left( \mathbf{\xi }\right) \sqrt{A_{1}\left( \mathbf{\xi ,x}%
\right) }\left( 1-\frac{1}{2}\left( V_{1}\left( \mathbf{\xi },\mathbf{y}%
_{2}\right) +V_{1}\left( \mathbf{\xi },\mathbf{y}_{1}\right) \right) \right) 
}{A_{1}\left( \mathbf{\xi ,\xi }\right) \sqrt{A_{1}\left( \mathbf{x},\mathbf{%
x}\right) }\sqrt{\left( \mathbf{x}-\mathbf{\xi }\right) ^{2}}}d\mathbf{\xi }
\nonumber \\
&&\times \left( V_{1}\left( \mathbf{\xi },\mathbf{y}_{2}\right) -V_{1}\left( 
\mathbf{\xi },\mathbf{y}_{1}\right) \right) r  \label{a6.8}
\end{eqnarray}%
where%
\[
r=\frac{\sqrt{\left( \mathbf{x-\xi }\right) ^{2}}}{\sqrt{A_{1}\left( \mathbf{%
\xi },\mathbf{x}\right) }} 
\]%
\begin{equation}
B_{2}\left( \mathbf{y}_{1},\mathbf{y}_{2}\right) =-2\frac{G}{c^{2}}%
\int_{V}\rho _{0}\left( \mathbf{\xi }\right) \left( V_{1}\left( \mathbf{\xi }%
,\mathbf{y}_{2}\right) -V_{1}\left( \mathbf{\xi },\mathbf{y}_{1}\right)
\right) d\mathbf{\xi +}\mathcal{O}\left( \varepsilon ^{3}\right)
\label{a6.9}
\end{equation}%
\[
\delta C_{2}\left( \mathbf{y}_{1},\mathbf{y}_{2}\right) =-\frac{G}{c^{2}}%
\int_{V}\frac{\rho _{0}\left( \mathbf{\xi }\right) \sqrt{A_{1}\left( \mathbf{%
\xi ,x}\right) }\left( \left( V_{1}\left( \mathbf{\xi },\mathbf{y}%
_{2}\right) -V_{1}\left( \mathbf{\xi },\mathbf{y}_{1}\right) \right) \frac{%
\sqrt{\left( \mathbf{x-\xi }\right) ^{2}}}{\sqrt{A_{1}\left( \mathbf{\xi },%
\mathbf{x}\right) }}\right) ^{2}}{A_{1}\left( \mathbf{\xi ,\xi }\right) 
\sqrt{A_{1}\left( \mathbf{x},\mathbf{x}\right) }\sqrt{\left( \left( \mathbf{x%
}-\mathbf{\xi }\right) ^{2}\right) }}d\mathbf{\xi } 
\]%
\begin{equation}
=-\frac{G}{c^{2}}\int_{V}\frac{\rho _{0}\left( \mathbf{\xi }\right) \sqrt{%
\left( \mathbf{x-\xi }\right) ^{2}}\left( V_{1}\left( \mathbf{\xi },\mathbf{y%
}_{2}\right) -V_{1}\left( \mathbf{\xi },\mathbf{y}_{1}\right) \right) ^{2}}{%
A_{1}\left( \mathbf{\xi ,\xi }\right) \sqrt{A_{1}\left( \mathbf{x},\mathbf{x}%
\right) A_{1}\left( \mathbf{\xi },\mathbf{x}\right) }}d\mathbf{\xi =}%
\mathcal{O}\left( \varepsilon ^{3}\right)  \label{a6.10}
\end{equation}%
We obtain%
\begin{equation}
V_{2}\left( \mathbf{y}_{1},\mathbf{y}_{2}\right) =\frac{2G}{c^{2}}\int_{V}%
\frac{\rho _{0}\left( \mathbf{\xi }\right) \sqrt{A_{1}\left( \mathbf{\xi ,x}%
\right) }\left( 1-\frac{1}{2}\left( V_{1}\left( \mathbf{\xi },\mathbf{y}%
_{2}\right) +V_{1}\left( \mathbf{\xi },\mathbf{y}_{1}\right) \right) \right)
^{2}}{A_{1}\left( \mathbf{\xi ,\xi }\right) \sqrt{A_{1}\left( \mathbf{x},%
\mathbf{x}\right) }\sqrt{\left( \mathbf{x}-\mathbf{\xi }\right) ^{2}}}d%
\mathbf{\xi +}\mathcal{O}\left( \varepsilon ^{3}\right)  \label{a6.11}
\end{equation}%
\begin{eqnarray}
V_{2}\left( \mathbf{y}_{1},\mathbf{y}_{2}\right) &=&V_{1}\left( \mathbf{y}%
_{1},\mathbf{y}_{2}\right) +\frac{G}{c^{2}}\int_{V}\frac{\rho _{0}\left( 
\mathbf{\xi }\right) \left( -V_{1}\left( \mathbf{\xi ,x}\right)
+2V_{1}\left( \mathbf{\xi ,\xi }\right) +V_{1}\left( \mathbf{x},\mathbf{x}%
\right) \right) }{\sqrt{\left( \mathbf{x}-\mathbf{\xi }\right) ^{2}}}d%
\mathbf{\xi }  \nonumber \\
&&-\frac{G}{c^{2}}\int_{V}\frac{\rho _{0}\left( \mathbf{\xi }\right) \left(
V_{1}\left( \mathbf{\xi },\mathbf{y}_{2}\right) +V_{1}\left( \mathbf{\xi },%
\mathbf{y}_{1}\right) \right) }{\sqrt{\left( \mathbf{x}-\mathbf{\xi }\right)
^{2}}}d\mathbf{\xi +}\mathcal{O}\left( \varepsilon ^{3}\right)  \label{a6.12}
\end{eqnarray}%
\begin{eqnarray}
V_{2}\left( \mathbf{y}_{1},\mathbf{y}_{2}\right) &=&V_{1}\left( \mathbf{y}%
_{1},\mathbf{y}_{2}\right) +\frac{2G}{c^{2}}\int_{V}\frac{\rho _{0}\left( 
\mathbf{\xi }\right) V_{1}\left( \mathbf{\xi ,\xi }\right) }{\sqrt{\left( 
\mathbf{x}-\mathbf{\xi }\right) ^{2}}}d\mathbf{\xi }+\mathcal{O}\left(
\varepsilon ^{3}\right)  \label{a6.14} \\
&&+\frac{G}{c^{2}}\int_{V}\frac{\rho _{0}\left( \mathbf{\xi }\right) \left(
-V_{1}\left( \mathbf{\xi ,x}\right) +V_{1}\left( \mathbf{x},\mathbf{x}%
\right) -V_{1}\left( \mathbf{\xi },\mathbf{y}_{2}\right) -V_{1}\left( 
\mathbf{\xi },\mathbf{y}_{1}\right) \right) }{\sqrt{\left( \mathbf{x}-%
\mathbf{\xi }\right) ^{2}}}d\mathbf{\xi }  \nonumber
\end{eqnarray}

Estimation of (\ref{a6.14}) in the case, when $\left\vert \mathbf{y}%
_{1}\right\vert ,\left\vert \mathbf{y}_{2}\right\vert ,\left\vert \mathbf{x}%
\right\vert \gg R$, has the form%
\begin{equation}
V_{2}\left( \mathbf{y}_{1},\mathbf{y}_{2}\right) =V_{1}\left( \mathbf{y}_{1},%
\mathbf{y}_{2}\right) +\frac{6}{5}\varepsilon ^{2}\frac{R}{\left\vert 
\mathbf{x}\right\vert }-\frac{\varepsilon ^{2}}{2}\frac{R^{2}}{\left\vert 
\mathbf{x}\right\vert ^{2}}\left( 1+\frac{2\left\vert \mathbf{x}\right\vert 
}{\left\vert \mathbf{y}_{1}\right\vert }+\frac{2\left\vert \mathbf{x}%
\right\vert }{\left\vert \mathbf{y}_{2}\right\vert }\right) +\mathcal{O}%
\left( \varepsilon ^{3}\right)  \label{a6.15}
\end{equation}%
where $V_{1}\left( \mathbf{y}_{1},\mathbf{y}_{2}\right) $ is determined by
the relation (\ref{a6.6}), and%
\begin{equation}
\varepsilon =\frac{2GM}{c^{2}R}\ll 1  \label{a6.15a}
\end{equation}%
In the case, when $\mathbf{y}_{1}=\mathbf{y}_{2}=\mathbf{x}$,\textbf{\ }we
have%
\begin{equation}
V_{2}\left( \mathbf{x},\mathbf{x}\right) =V_{1}\left( \mathbf{x},\mathbf{x}%
\right) +\frac{6}{5}\varepsilon ^{2}\frac{R}{\left\vert \mathbf{x}%
\right\vert }-\frac{5}{2}\varepsilon ^{2}\frac{R^{2}}{\left\vert \mathbf{x}%
\right\vert ^{2}}+\mathcal{O}\left( \varepsilon ^{3}\right)  \label{a6.16b}
\end{equation}

We obtain for the quantity $B_{2}\left( \mathbf{y}_{1},\mathbf{y}_{2}\right) 
$ for $\left\vert \mathbf{y}_{2}\right\vert ,\left\vert \mathbf{y}%
_{1}\right\vert \gg R$ 
\begin{eqnarray}
B_{2}\left( \mathbf{y}_{1},\mathbf{y}_{2}\right) &=&-2\frac{G}{c^{2}}%
\int_{V}\rho _{0}\left( \mathbf{\xi }\right) \left( V_{1}\left( \mathbf{\xi }%
,\mathbf{y}_{2}\right) -V_{1}\left( \mathbf{\xi },\mathbf{y}_{1}\right)
\right) d\mathbf{\xi +}\mathcal{O}\left( \varepsilon ^{3}\right)  \nonumber
\\
&=&-2\frac{GM}{c^{2}}\left( V_{1}\left( 0,\mathbf{y}_{2}\right) -V_{1}\left(
0,\mathbf{y}_{1}\right) \right) \mathbf{+}\mathcal{O}\left( \varepsilon
^{3}\right)  \nonumber \\
&=&-\varepsilon ^{2}R^{2}\left( \frac{1}{\left\vert \mathbf{y}%
_{2}\right\vert }-\frac{1}{\left\vert \mathbf{y}_{1}\right\vert }\right) 
\mathbf{+}\mathcal{O}\left( \varepsilon ^{3}\right)  \label{a6.16}
\end{eqnarray}%
\begin{equation}
B_{2}\left( \mathbf{x},\mathbf{x}\right) =0  \label{a6.17}
\end{equation}

Thus, for small $\varepsilon =2GM/\left( Rc^{2}\right) $ and $\left\vert 
\mathbf{x}\right\vert \gg R$, the calculated value of metric tensor,
determined by the quantities $V_{1}\left( \mathbf{y}_{1},\mathbf{y}%
_{1}\right) ,B_{1}\left( \mathbf{y}_{1},\mathbf{y}_{1}\right) ,\delta
C_{1}\left( \mathbf{y}_{1},\mathbf{y}_{1}\right) $ coincides with the metric
tensor, calculated in Newtonian approximation of the general relativity.

At large values of parameter $\varepsilon $ the quantity $V\left( \mathbf{x,x%
}\right) $ remains to be less, than unity. Indeed, setting $\mathbf{y}_{1}=%
\mathbf{y}_{2}=\mathbf{x}$ in exact equations (\ref{a5.36}) - (\ref{a5.38}),
we obtain 
\begin{equation}
V\left( \mathbf{x},\mathbf{x}\right) =\frac{2G}{c^{2}}\int_{V}\frac{\rho
\left( \mathbf{\xi }\right) A\left( \mathbf{\xi ,x}\right) \left( 1-\frac{1}{%
2}\left( V\left( \mathbf{\xi },\mathbf{x}\right) +V\left( \mathbf{\xi },%
\mathbf{x}\right) \right) \right) ^{2}}{A\left( \mathbf{\xi ,\xi }\right) 
\sqrt{A\left( \mathbf{x},\mathbf{x}\right) }\sqrt{B^{2}\left( \mathbf{\xi },%
\mathbf{x}\right) +A\left( \mathbf{\xi },\mathbf{x}\right) \left( \left( 
\mathbf{x}-\mathbf{\xi }\right) ^{2}-2\delta C\left( \mathbf{\xi ,x}\right)
\right) }}d\mathbf{\xi }  \label{a6.18}
\end{equation}%
\begin{equation}
B\left( \mathbf{x},\mathbf{x}\right) =0,\qquad \delta C\left( \mathbf{x},%
\mathbf{x}\right) =0  \nonumber
\end{equation}%
Rewriting equation (\ref{a6.18}) in the form%
\[
V\left( \mathbf{x},\mathbf{x}\right) \sqrt{1-V\left( \mathbf{x},\mathbf{x}%
\right) } 
\]%
\begin{equation}
=\frac{2G}{c^{2}}\int_{V}\frac{\rho \left( \mathbf{\xi }\right) A\left( 
\mathbf{\xi ,x}\right) \left( 1-\frac{1}{2}\left( V\left( \mathbf{\xi },%
\mathbf{x}\right) +V\left( \mathbf{\xi },\mathbf{x}\right) \right) \right)
^{2}}{A\left( \mathbf{\xi ,\xi }\right) \sqrt{B^{2}\left( \mathbf{\xi },%
\mathbf{x}\right) +A\left( \mathbf{\xi },\mathbf{x}\right) \left( \mathbf{x}-%
\mathbf{\xi }\right) ^{2}-2A\left( \mathbf{\xi },\mathbf{x}\right) \delta
C\left( \mathbf{\xi ,x}\right) }}d\mathbf{\xi }  \label{a6.19}
\end{equation}%
we conclude, that equation (\ref{a6.19}) contains only solutions with $%
V\left( \mathbf{x},\mathbf{x}\right) \leq 1$. In other words, component $%
g_{00}=c^{2}\left( 1-V\left( \mathbf{x,x}\right) \right) $ of the metric
tensor cannot change its sign. \textit{It means that non-rotating physical
body of any size and of any mass cannot generate a black hole}.

This result disagrees with the result of general relativity, but it agrees
with the common sense. To obtain the reason of such unexpected result, we
calculate the quantities $A,B,\delta C$ inside the uniform heavy sphere of
radius $R$ and mass $M$. At calculation we suppose that the quantity 
\begin{equation}
\varepsilon =\frac{r_{g}}{R}=\frac{2GM}{c^{2}R}\ll 1  \label{a6.20}
\end{equation}%
where $r_{g}$ is the gravitational radius of the sphere.

For $\left\vert \mathbf{x}\right\vert <R$ results of calculations looks as
follows (Details of calculations are rather bulky, and we omit them)%
\begin{equation}
V_{2}\left( \mathbf{x},\mathbf{x}\right) =\varepsilon \left( \frac{3}{2}-%
\frac{1}{2}\frac{\mathbf{x}^{2}}{R^{2}}\right) -\varepsilon ^{2}\frac{153}{64%
}+\varepsilon ^{2}\frac{37}{32}\frac{\mathbf{x}^{2}}{R^{2}}-\varepsilon ^{2}%
\frac{61}{320}\frac{\left\vert \mathbf{x}\right\vert ^{4}}{R^{4}}\mathbf{+}%
\mathcal{O}\left( \varepsilon ^{3}\right)  \label{a6.21}
\end{equation}%
The gravitational force inside the region $\left\vert \mathbf{x}\right\vert
<R$ has the form 
\begin{equation}
\mathbf{F}=\mathbf{\nabla }V_{2}\left( \mathbf{x},\mathbf{x}\right) =-\frac{%
\varepsilon }{R^{2}}\mathbf{x+}\frac{\varepsilon ^{2}}{R^{2}}\frac{37}{16}%
\mathbf{x-}\frac{61}{80}\frac{\varepsilon ^{2}}{R^{2}}\frac{\left\vert 
\mathbf{x}\right\vert ^{2}}{R^{2}}\mathbf{x,\qquad }\left\vert \mathbf{x}%
\right\vert <R  \label{a6.22}
\end{equation}%
It follows from (\ref{a6.22}), that if $\varepsilon >\frac{16}{37}\approx
0.43$, the region, where the gravitational force is directed from the
center, appears near the point $\mathbf{x}=0$. If $\varepsilon \geq 0.65$,
the gravitational force is directed from the center of the sphere in the
whole region $\left\vert \mathbf{x}\right\vert <R$.

Thus, inside the heavy sphere the regions of antigravitation may appear at
large values of $\varepsilon $. To understand this unexpected circumstance,
let us note, that dynamical (not completely relativistic) approach and
geometrical (completely relativistic) approach to gravitational phenomena
disagree in some points.

The Newtonian gravitational potential of a uniform heavy sphere of radius $R$
has the form 
\begin{equation}
\varphi \left( \mathbf{x}\right) =\left\{ 
\begin{array}{ccc}
\frac{GM}{\left\vert \mathbf{x}\right\vert } & \text{if} & \left\vert 
\mathbf{x}\right\vert >R \\ 
\frac{3GM}{2R}-\frac{GM}{2R^{3}}\left\vert \mathbf{x}\right\vert ^{2} & 
\text{if} & \left\vert \mathbf{x}\right\vert <R%
\end{array}%
\right.  \label{a6.23}
\end{equation}%
where $M$ is the of the sphere. The gravitational potential $\varphi $ is
maximal at the point $\mathbf{x}=0$, whereas the gravitational force $%
\mathbf{F}=\mathbf{\nabla }\varphi $ is minimal at the point $\mathbf{x}=0$ (%
$\mathbf{F}=0$ at $\mathbf{x}=0$). The space-time geometry is connected with
the gravitational potential $g_{00}=\left( c^{2}-2\varphi \right) $, but not
with the gravitational force $\mathbf{F}$.

Gravitational potential $\varphi $ inside the hallow sphere of mass $M$ is
proportional to the mass $M$, but $\varphi =$const, and the gravitational
force $\mathbf{F}=0$ inside the sphere. From dynamic (differential)
viewpoint this fact is explained as a result compensation of gravitational
influence of different parts of the hallow sphere. If the gravitational law
distinguishes from the Newtonian one, such a compensation may disappear, and
an induced antigravitation may appear, because the attraction force,
generated by any part of the sphere, is directed from the center of the
sphere.

From the geometric (integral) viewpoint an appearance of the induced
antigravitation regions is natural, because the gravitational potential
increases in such regions with increase of amount of the matter. As to the
gravitational force, it may have any direction.

\section{Concluding remarks}

Thus, the extended general relativity (EGR) is the next stage of the physics
geometrization. At this stage we have the monistic conception, containing
only one fundamental quantity: world function $\sigma $. The gravitational
field, which is one of fundamental quantities of the general relativity
(GR), is now only an attribute of the world function. From viewpoint of
extended general relativity the gravitational field is not a physical
essence. It is only a manner of the particle interaction description. In
particular, from viewpoint of EGR the gravitational field cannot exist
separate from the matter. Such a change of approach to the gravitational
field is connected with a usage of the relativistic concept of the events
nearness.

Any monistic conception is a result of development of the preceding
pluralistic conception, and the monistic conception is more perfect as a
rule, than the preceding one. The extended general relativity (EGR) is
obtained as a result of overcoming of defects of the general relativity
(GR): (1) usage of only inconsistent Riemannian space-time geometry, (2) use
of inadequate (nonrelativistic) concepts and quantities. EGR is to be
considered as a more perfect conception, than GR. Results obtained in the
framework of EGR are more dependable, than results, obtained in the
framework of GR. In particular, conclusion on impossibility of the dark hole
existence in EGR is more dependable, than existence of the black holes in
the framework of GR. Besides, impossibility of the gravitational collapsing,
leading to a formation of a black hole , is confirmed by appearance of
induced antigravitation in EGR.

Besides, the mathematical technique of EGR is the same for all (continuous
and discrete) geometries. Dynamic equations for the world function are
written in the coordinateless form. This circumstance admits one to
eliminate consideration of any coordinate transformation.

There is a possibility, that some problems of contemporary cosmology (dark
energy, dark energy) are a result of imperfect theory of gravitation. More
correct results of EGR, concerning dark holes, admit to hope, that EGR will
be able to solve difficult problems of contemporary cosmology.


\begin{thebibliography}{99}
\bibitem{R2001} Yu.A.Rylov, Geometry without topology as a new conception of
geometry. \textit{Int. Jour. Mat. \& Mat. Sci.} \textbf{30}, iss. 12,
733-760, (2002),

\bibitem{R2007} Yu.A.Rylov, Non-Euclidean method of the generalized geometry
construction and its application to space-time geometry in \textit{Pure and
Applied Differential geometry} pp.238-246. eds. Franki Dillen and Ignace Van
de Woestyne. Shaker Verlag, Aachen, 2007. Available also at \textit{%
http://arXiv.org/abs/Math.GM/0702552}

\bibitem{S60} J.L.Synge, \textit{Relativity: the General Theory. }Amsterdam,
North-Holland Publishing Company, 1960.

\bibitem{R91} Yu.A.Rylov, Non-Riemannian model of the space-time responsible
for quantum effects. \textit{Journ. Math. Phys}. \textbf{32(8)}, 2092-2098,
(1991).

\bibitem{R2010} Yu. A. Rylov, Logical reloading as overcoming of crisis in
geometry. \textit{e-print} \textit{1005.2074}

\bibitem{R2010a} Yu. A. Rylov, Monistic conception of geometry. \textit{%
e-print 1009.2815}

\bibitem{R2008} Yu.A.Rylov, Generalization of relativistic particle dynamics
on the case of non-Riemannian space-time geometry.. \textit{Concepts of
Physics} \textbf{6}, Number.4, p 605, (2009). ISSN1897-2357 See also \textit{%
e-print, http://arXiv.org/abs/0811.4562}

\bibitem{V2005} Yu.S.Vladimirov, \textit{Geometrodynamics}. Moscow, Binom,
Laboratory of sciences, 2005. chp.14.

\bibitem{F1929} A.D.Fokker, Ein invarianter Variationssatz f\"{u}r die
Bewegung mehrer elektrischer Massenteilchen. \textit{Z.Phys.} Bd. \textbf{%
58, }386-393, (1929).

\bibitem{F55} V.A. Fock, \textit{Theory of space, time and gravitation},
GITTL, Moscow, 1955. (in Russian) Sec. 53,54.

\bibitem{R2007a} Yu.A.Rylov, Different conceptions of Euclidean geometry
http://arXiv.org/abs/0709.2755.%
\[
\]
\end{thebibliography}
\end{document}